\documentclass[12pt,journal,onecolumn,draftclsnofoot]{IEEEtran}
\pdfoutput=1
\usepackage{color}
\usepackage{graphicx}
\usepackage{epstopdf}
\usepackage{subfigure}
\usepackage{amsmath}
\usepackage{amsfonts}
\usepackage{algorithmicx}
\usepackage{algorithm}
\usepackage[noend]{algpseudocode}
\usepackage{cite}
\usepackage{url}
\usepackage[export]{adjustbox}
\usepackage[utf8]{inputenc}
\usepackage[font=small]{caption}
\usepackage{cases}
\usepackage{amsmath}
\usepackage{multirow}
\usepackage{array}
\usepackage{siunitx}
\usepackage{xspace}
\usepackage{fancyhdr}
\usepackage{wrapfig}

\newcommand{\figSize}{0.31\columnwidth} %

\newtheorem{defi}{Definition}

\newtheorem{thm}{Theorem}

\definecolor{mybrown}{RGB}{128,0,0}

\newcommand{\comment}[1]{}

\newcommand{\name}{Panda\xspace}
\DeclareMathOperator*{\E}{\mathbb{E}}

\begin{document}
\title{\name: Neighbor Discovery on\\ a Power Harvesting Budget\vspace*{-.10349in}}
\author{\IEEEauthorblockN{Robert Margolies\IEEEauthorrefmark{1},
  Guy Grebla\IEEEauthorrefmark{1},
  Tingjun Chen\IEEEauthorrefmark{1},
  Dan Rubenstein\IEEEauthorrefmark{2},
  Gil Zussman\IEEEauthorrefmark{1}}
 \IEEEauthorblockA{\IEEEauthorrefmark{1}Electrical Engineering and \IEEEauthorrefmark{2}Computer Science, Columbia University \\
\{robm, guy, tingjun, gil\}@ee.columbia.edu, danr@cs.columbia.edu}\thanks{A partial and preliminary version of this paper will appear in IEEE INFOCOM'16~\cite{Margolies_INFOCOM16}, April 2016.}
\thanks{This research was supported by NSF grants CCF-09-64497 and CNS-10-54856, and the People Programme (Marie Curie Actions) of the European Union's Seventh Framework Programme (FP7/2007-2013) under REA grant agreement no.~[PIIF-GA-2013-629740].11}}

\maketitle

\begin{abstract}
Object tracking applications are gaining popularity and will soon utilize Energy Harvesting (EH) low-power nodes that will consume power mostly for Neighbor Discovery (ND) (i.e., \emph{identifying} nodes within communication range). Although ND protocols were developed for sensor networks, the \emph{challenges posed by emerging EH low-power transceivers were not addressed}. Therefore, we \emph{design an ND protocol tailored for the characteristics of a representative EH prototype}: the TI eZ430-RF2500-SEH.
We present a generalized model of ND accounting for unique prototype characteristics (i.e., energy costs for transmission/reception, and transceiver state switching times/costs). Then, we present the Power Aware Neighbor Discovery Asynchronously (\name) protocol in which nodes transition between the sleep, receive, and transmit states.
We analyze \name and select its parameters to maximize the \emph{ND rate} subject to a homogenous power budget. We also present \name-D, designed for {\em non-homogeneous} EH nodes. We perform extensive testbed evaluations using the prototypes and study various design tradeoffs. We demonstrate a small difference (less then 2\%) between experimental and analytical results, thereby confirming the modeling assumptions. Moreover, we show that \name improves the ND rate by up to 3x compared to related protocols. Finally, we show that \name-D operates well under non-homogeneous power harvesting.
\end{abstract}


%

\section{Introduction}
\label{sec:intro}

Object tracking and monitoring applications are gaining popularity within the realm of the Internet-of-Things~\cite{Atzori20102787_IoT}. Emerging low-power wireless nodes that can be attached to physical objects are enablers for such applications. Often, these nodes are meant to interact with a reader, but architectures are emerging that handle scenarios where no reader may be present, or where the number of nodes overwhelms the readers' availability. These scenarios can be supported by Energy Harvesting (EH) tags (e.g., \cite{Ulukus_JSACreview, Margolies_EnHANTS_TOSN} and references therein) that are able to communicate peer-to-peer and are powered by an ambient energy source (e.g., light).

\begin{figure}[t]
\centering
\includegraphics[width = 0.6\columnwidth]{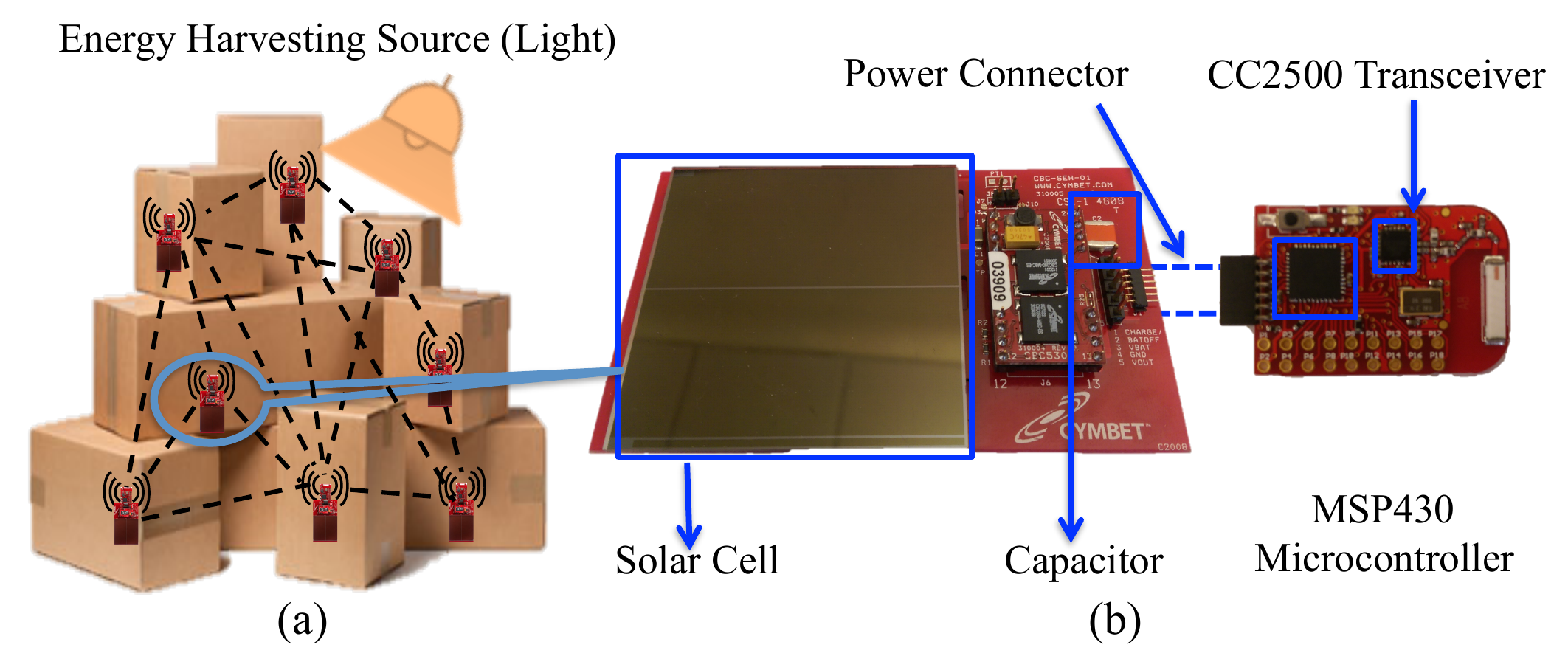}
\vspace*{-.1in}
\caption{(a) EH nodes can be attached to everyday objects (e.g., boxes) and utilize neighbor discovery protocols for inventory tracking and monitoring applications.\label{fig:intro} (b) The prototype EH node stores harvested energy in a capacitor and powers the microcontroller and transceiver. \label{fig:EHBprototype}}
\end{figure}

Such EH nodes will enable tracking applications in healthcare, smart buildings, assisted living, manufacturing, supply chain management, and intelligent transportation as discussed in \cite{Wang_SigComm2013,liu2013ambient}. 
An example application, illustrated in Fig.~\ref{fig:intro}(a), is a large warehouse that contains many inventory items, each of which is equipped with an EH node. 
Each node has an ID that corresponds to the physical object (item). The nodes utilize a Neighbor Discovery (ND) protocol to \emph{identify} neighbors which are within  communication range, and therefore, the system can collect information about the objects' whereabouts. A simple application is identifying misplaced objects: often when an item is misplaced (e.g., in a furniture warehouse, a box of table parts is moved to an area with boxes of bed parts), its ID is significantly different from the IDs of its neighbors. In such a case, the misplaced node can, for instance, flash a low-power LED to indicate that it is lost.




In this paper, {\em we develop an ND protocol for Commercial Off-The-Shelf (COTS) EH nodes, based on the TI eZ430-RF2500-SEH} \cite{TIehavkit} (shown in Fig.~\ref{fig:EHBprototype}(b)). The nodes harvest ambient light to supply energy to a low-power microcontroller and transceiver. To maintain \emph{perpetual} tracking of the (potentially) mobile objects, ND must be run continuously with the node operating in an \emph{ultra-low-power mode} that consumes power at the rate of power harvested\cite{Gorlatova_NetworkingLowPower}. 
Our objective is {\bf to maximize the rate in which nodes {\em discover} their neighbors, given a constrained power budget at each EH node.}

ND has always been an important part of many network scenarios~\cite{radi2014network,Sun_NDSurvey}.
Yet, to consume power at the rate of power harvested, EH nodes require \emph{extremely limited power budgets}: we show that, even with optimized power spending, the duty cycles are between 0.1-0.6\%.
Therefore, numerous assumptions from related works (e.g., \cite{Bakht_mobicom2012,sun2014hello}) no longer hold, including that switching times (between the sleep, receive, and transmit states) draw negligible power and that the power costs to send and receive are identical (see Section~\ref{sec:related} for details). 
Furthermore, in the envisioned applications, the node's main task is to perform ND, 
and thus, the power consumed by ND is the dominant component of the power budget.

Hence, we 
design, analyze, and experiment with {\bf \name -- Power Aware Neighbor Discovery Asynchronously},\footnote{The protocol name, Panda, relates to the animal as both EH nodes and Pandas spend the majority of their time sleeping to conserve energy.} an ND protocol that {\em maximizes the  average discovery rate under a given power budget}. 
 The main contributions of this paper are:



{\noindent\bf (C1) Radio Characterization:} We model a generic ultra-low-power EH node that captures the capabilities of our prototype (Fig.~\ref{fig:EHBprototype}(b)). We also study, for the first time, important properties of the radio in the context of ND.
We show that characteristics such as the power consumption and the time to transition between the different states (e.g., sleep to listen) are crucial to incorporate into the design of ND protocols for EH nodes.



{\noindent\bf (C2) \name Protocol}:
We develop the \name~protocol in which an EH node discovers its neighbors by transitioning between the sleep, receive, and transmit states at rates that satisfy a power budget. Furthermore, we present \name-Dynamic (\name-D), which extends \name's applicability to non-homogeneous power harvesting and multihop topologies.

{\noindent\bf (C3) Protocol Optimization:}
Using techniques from renewal theory, we derive closed form expressions for the discovery rate and the power consumption. 
We develop the 
\name Configuration Algorithm (PCA) to determine the node's duration in each state (sleep, receive, transmit), such that the discovery rate is maximized, while meeting the power budget. 
The solution obtained by the PCA is numerically shown to be between 94--99.9\% of the optimal for all scenarios considered.


{\noindent\bf (C4) Experimental Evaluation:}
Using TI eZ430-RF2500-SEH EH nodes \cite{TIehavkit}, 
we show that the real-life discovery rates are within 2\% of the analytically predicted values, demonstrating the practicality of our model. Moreover, we show that \name's \emph{experimental} discovery rate is up to 3 times higher than the discovery rates from simulations of two of the previously best known low-power ND protocols~\cite{Bakht_mobicom2012,McGlynn_mobihoc01}.
Furthermore, we demonstrate that \name-D adjusts the rate of ND for scenarios with non-homogenous power harvesting and multihop topologies. 

The rest of the paper is organized as follows. In Section~\ref{sec:related} we discuss related work. In Section~\ref{sec:node_model} we present the system model. In Sections~\ref{sec:protocol} and~\ref{sect:solutions}, we present and optimize \name, respectively. 
 In Section~\ref{sect:dynamic}, we present the \name-D protocol.
In Section~\ref{sect:experiments}, we evaluate \name~experimentally. We conclude in Section~\ref{sect:conc}.

\section{Related Work}
\label{sec:related}
\label{sect:related}

ND for low-power wireless networks is a well studied problem (see\cite{Atzori20102787_IoT,radi2014network,Sun_NDSurvey} for a summary).
The protocols can be categorized into deterministic
(e.g., \cite{Dutta_disco08,Bakht_mobicom2012,Kandhalu_uconnect10,sun2014hello,purohit_wiflock11})
and probabilistic (e.g.,~\cite{you_aloha_nd11, McGlynn_mobihoc01}).  
Deterministic protocols 
focus on guaranteeing an upper bound on \emph{discovery latency}, while the choice of parameters (e.g., prime numbers) is often limited.
On the other hand, the most well-known probabilistic protocol~\cite{McGlynn_mobihoc01}, has a better average ND rate, 
but suffers from an unbounded discovery latency.
Our probabilistic protocol, \name, is fundamentally different: other protocols (i) are constrained by a duty cycle, instead of a power budget, (ii) 
do not account for channel collisions (e.g., when two nodes transmit at the same time), 
(iii) rely on each node maintaining synchronized time slots,\footnote{It was shown in~\cite{Dutta_disco08} how the aligned time slot assumption can be relaxed. Yet, practical considerations such as selecting the slot duration and avoiding collisions are not described.} or (iv) do not consider practical hardware energy consumption costs (i.e., the power consumed by the radio to transition between different states).

To the best of our knowledge, {\em \name is the first ND protocol for EH nodes} and the \emph{first attempt to maximize the discovery rate, given a power budget}. As such, \name will operate with duty cycles between 0.1-0.6\%, which is an order of magnitude lower than those typically considered in prior works~\cite{Sun_NDSurvey}. 

In our experiments, we use hardware from~\cite{TIehavkit}. There are also numerous other hardware options for EH nodes~\cite{TanejaDesign,enocean,Margolies_EnHANTS_TOSN},
computational RFIDs \cite{Zhang2014}, and \mbox{mm$^3$-scale} wireless devices \cite{UMichMote}.  
Additionally, there are other radio features that achieve low energy consumption. For example, preamble-sampling and wake up radios were investigated in~\cite{el2004wisemac} and~\cite{dunkels2011contikimac}, respectively, for WSNs.
However, the added power consumption of these features makes them impractical for the EH nodes we study. Furthermore, numerous options for low-power wireless communication exist (e.g., Bluetooth Low Energy~\cite{ble}). However, \cite{TIehavkit} is one of the increasingly popular low-power EH nodes which seamlessly support wireless protocol development.

\section{System Model}
\label{sect:model}
\label{sec:node_model}
In this section, we describe our prototypes, 
based on which, we introduce the notation and the system model. 



\subsection{Prototype Description}
\label{sec:prototype}
The prototype is shown in Fig.~\ref{fig:EHBprototype}(b) and is based on the commercially available TI eZ430-RF2500-SEH \cite{TIehavkit}. We made some modifications to the hardware as summarized in
Table~\ref{tab:modifications}. 

\begin{table}[t!]
\centering
\scriptsize
\caption{Modifications to the TI eZ430-RF2500 \label{tab:modifications}}
\begin{tabular}{|p{0.15\columnwidth}|p{0.75\columnwidth}|}
\hline Component & Problem/Modification \\
\hline
\hline Energy Storage & On-board battery cannot be monitored; disable on-board battery and replace with an external capacitor. \\
\hline Solar Panel & On-board solar cell cannot be monitored; disable on-board cell, measure power harvested by connecting a ammeter in series with solar cell from~\protect\cite{sanyo_sc}. \\
\hline Power Consumption & Unable to track power consumed; measure consumed power with an oscilloscope across a 10\si{\ohm} sense resistor, placed in series with the transceiver and the microcontroller.  \\
\hline $12$\si{\kilo\hertz} Clock Source & Clock frequency varies by up to $20\%$ for each node; rectified by manually measuring/calibrating the number of clock ticks in one second for each device. \\
\hline
\end{tabular}
\end{table}

We now describe the prototype's components:

\noindent\textbf{Energy Harvesting Power Source: } The prototype harvests light from a Sanyo AM 1815 amorphous solar cell~\cite{sanyo_sc}. The solar cell is set to a fixed harvesting voltage of 1.02V\comment{Explain this in tech report} (no power point tracking techniques are used). To measure the power harvested, we place a ammeter   
in series with the solar cell.

\noindent\textbf{Energy Storage: }
The energy harvested by the solar cells is stored in a capacitor and the voltage is denoted by $V_{\rm cap}$. The voltage is regulated to 3.5\si{\volt} to power the node. We modified the board design to enable experimentation with varying capacitor sizes. Unless stated otherwise, we use a 30\si{\milli\farad} capacitor.
To ensure stable voltage regulation, 
a software cutoff is imposed; if $V_{\rm cap} \leq 3.6$\si{\volt}, the node enters and remains in a low-power sleep state until enough power is harvested such that $V_{\rm cap}$ exceeds the cutoff. 

\noindent \textbf{Low-Power Microcontroller:} A TI-MSP430 microcontroller~\cite{msp430} is used to provide computational capabilities. 
These include (i) sampling the capacitor voltage using an analog to digital converter (ADC), (ii) operating a low-power 12\si{\kilo\hertz} clock with an idle power draw of 1.6\si{\micro\watt} to instruct the node 
to enter and exit an ultra-low-power sleep state, 
and (iii) receiving and sending messages to the radio layer. 


\noindent\textbf{Low-Power Transceiver: } The prototype utilizes a CC2500 wireless transceiver (a 2.4\si{\giga\hertz} transceiver designed to provide low-power wireless communication)~\cite{CC2500} to send and receive messages. The transceiver operates at 250kbps and consumes $64.85$\si{\milli\watt} while in receive state. The transmission power can be set in software and we utilize levels between $-16$ and 1dBm, 
with a resulting power consumption between 53.25 and 86.82\si{\milli\watt}. At these levels, nodes within the same room typically have little or no packet loss.


\begin{figure}
\centering
\includegraphics[width = 0.6\columnwidth]{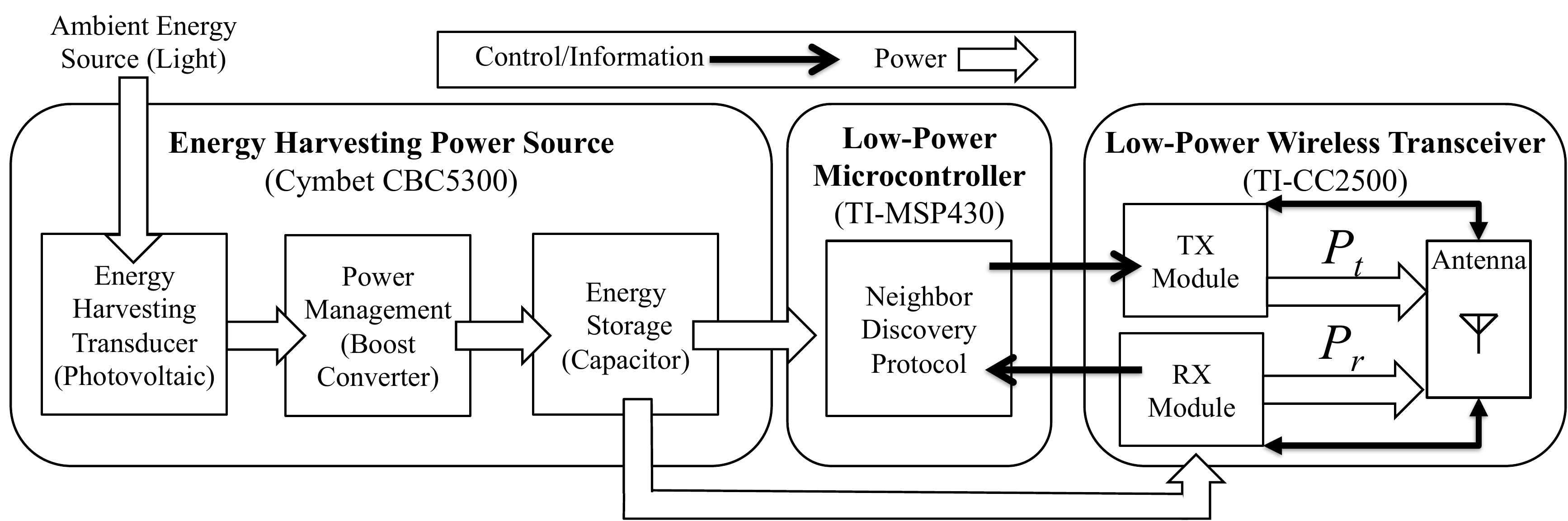}
\caption{A hardware model of the EH node (based on the TI eZ430-RF2500-SEH~\cite{TIehavkit}).\label{fig:sysmodel}}	
\end{figure}

\subsection{System Model} 
\label{sect:sys_model}
The model is based on the prototype and is shown pictorially in Fig.~\ref{fig:sysmodel}. Yet, it is generalizable to a class of other prototypes (e.g.,~\cite{Margolies_EnHANTS_TOSN}). A summary of the nomenclature from this point forward appears in Table~\ref{tbl:nomen}.


\begin{table}[t]
\centering
\caption{Nomenclature\label{tbl:nomen}}
\scriptsize
\begin{tabular}{|l|p{6.0cm}|l|p{6.0cm}|}
\hline {\bf Symbol} & {\bf Description} & {\bf Symbol} & {\bf Description} \\
\hline $V_{\rm cap}$ & The voltage of the capacitor (\si{\volt}) & $l$ & The duration of the listening period (\si{\milli\second})\\
\hline $N$ & Number of nodes & $\rho$ & Expected renewal duration (\si{\milli\second})\\
\hline $P_b$ & Average power spending budget (\si{\milli\watt}) & $Y$ & Denotes role of node in the renewal\\
\hline $P_t$ & Transmitting power consumption (\si{\milli\watt}) & $\eta()$ & Expected energy spending (\si{\micro\joule})\\
\hline $P_r$ & Listening/Receiving power consumption (\si{\milli\watt}) & $\Phi()$ & Expected power spending in a renewal (\si{\milli\watt})\\
\hline $C_{ij}$ & Energy cost to transition from state $i$ to $j$ (\si{\micro\joule}) & $\chi$ & Expected duration of idle listening (\si{\milli\second})\\
\hline $M$ & Discovery-packet duration (\si{\milli\second}) & $U$ & Discovery rate (\si{second}$^{-1}$)\\
\hline $\lambda$ & Rate of exponential distribution (\si{\milli\second}$^{-1}$) & & \\
\hline
\end{tabular}
\end{table}

A node can be in one out of three states, denoted by the set $\mathcal{S}=\{s,r,t\}$ for sleep ($s$), receive\footnote{We refer to the \emph{receive} and the \emph{listen} states synonymously as the power consumption of the prototype in both states is similar.} ($r$), and transmit ($t$). A node in state $i\in\mathcal{S}$ consumes power of $P_i$. Since the power consumption in sleep state is negligible, we assume $P_s=0$ throughout the paper and remark that all results can be easily applied for $P_s>0$, as described in Appendix~\ref{app:sleeping}.
For the power budgets we consider, the energy consumed by the radio to transition between different states is non-negligible. Hence, we denote by $C_{ij}$ the energy (\si{\micro\joule}) consumed to switch from state $i$ to state $j$ ($i,j\in \mathcal{S}$). 


Unfortunately, the prototype does not have explicit power awareness (unlike, e.g.,~\cite{Margolies_EnHANTS_TOSN}). Therefore, we impose a power budget, $P_b$ (\si{\milli\watt}) on each node. The power budget is set such that \emph{energy neutrality} is achieved: nodes consume power (on average) at the power harvesting rate~\cite{vigorito2007adaptive}. Hence, for an EH node harvesting more power (e.g., brighter light source), $P_b$ is higher.


We denote by $N$ the number of nodes in the network and present two important definitions:
\begin{defi}\label{defi:disc}
The {\bf discovery message} is a broadcast packet containing the ID of the transmitter.\footnote{In practice, the discovery message may include information on already discovered neighbors, thus enabling \emph{indirect} discoveries. However, we do not consider these indirect discoveries.}  A {\bf discovery} occurs when a node receives a discovery message from a neighbor. Multiple discoveries can occur per discovery message transmission.
\end{defi}
\begin{defi}\label{defi:discrate}
  The {\bf discovery rate}, denoted by $U$, is the expected number of discoveries in the network
  per second.
\end{defi}

The objective of the ND protocol is to \emph{maximize the discovery rate, subject to a given power budget}. This is in contrast to other works which seek to minimize the worst case discovery latency~\cite{Bakht_mobicom2012,sun2014hello}, subject to a duty cycle. As such, in Section~\ref{sect:experiments}, we also consider the discovery latency, or time in between discoveries, as a secondary performance metric.

\section{The \name~Protocol}
\label{sec:protocol}

In this section we describe and analyze \name, an asynchronous ND
protocol, which operates under a power budget. 



\subsection{Protocol Description}
\begin{figure}
\begin{minipage}{0.4\columnwidth}
\centering
\vspace*{0.12in}
{\includegraphics[width = 0.9\columnwidth]{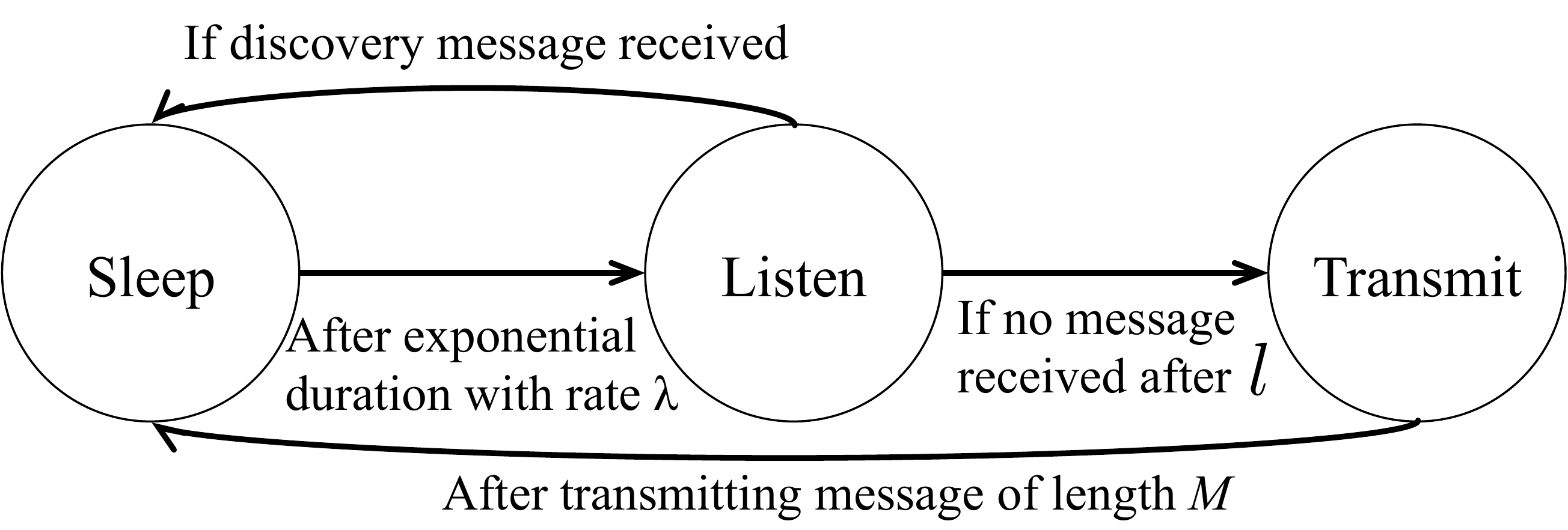}}
\vspace*{0.25in}
\caption{\name~protocol outline: EH nodes transition between radio states (sleep, listen, and transmit) to maintain within a power budget.\label{fig:nodeRen}}
\end{minipage}
\hspace*{0.03\columnwidth}
\begin{minipage}{0.57\columnwidth}
\centering
\includegraphics[width = 0.8\columnwidth]{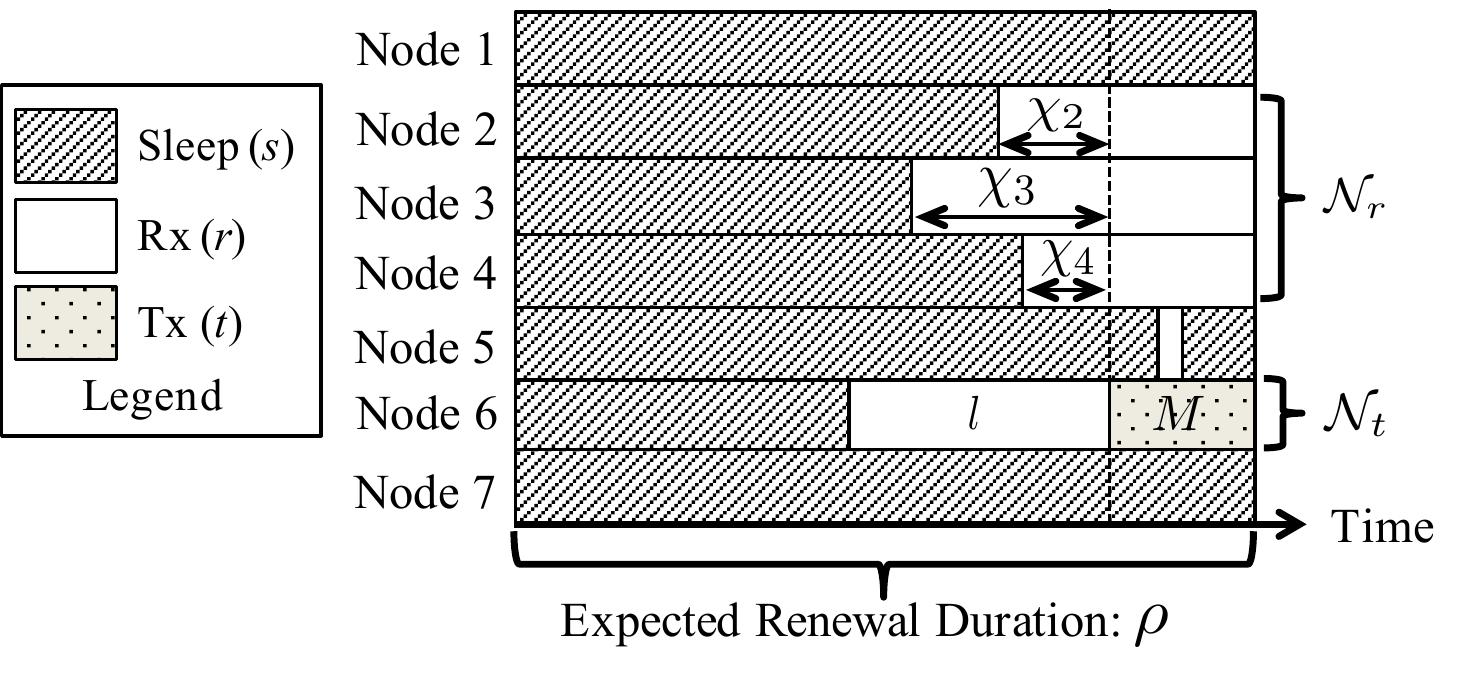}
\vspace*{-.15in}
\caption{Renewal process representing one renewal cycle for $N=7$ nodes: all nodes begin in a memoryless sleep state and the renewal restarts after the first node to wake up completes its transmission.\label{fig:netRen}}	
\end{minipage}
\end{figure}

Fig.~\ref{fig:nodeRen} depicts the state transition diagram for the \name~protocol, from sleep to listen to transmit and then back to sleep.
 To ensure perpetual operation under the power budget $P_b$ (\si{\milli\watt}), nodes initialize in a low-power sleep state to conserve energy. To maximize the discovery rate, \name follows a probabilistic approach in which nodes sleep for an 
exponential duration with rate $\lambda$ (\si{\milli\second}$^{-1}$). The probabilistic sleep duration prevents unwanted synchronization among subsets of nodes. 

Following sleep, nodes awaken and listen to the channel for discovery messages
from their neighbors for a fixed duration of $l$ (\si{\milli\second}). If a message is received, the node remains in the listen state until it completes reception of this message.
If no transmission is heard while in the listen state, the node transmits its
discovery message of fixed duration \mbox{$M$(\si{\milli\second}).}\footnote{The discovery message duration, $M$, is fixed, stemming from the fixed size of the node ID contained in the message.}

Note that in \name, similar to CSMA, nodes always listen before they transmit, and therefore, there are no packet collisions between two nodes in wireless communication range of one another. 
Additionally, after a message is transmitted, the node returns to the sleep state. Hence, there is no acknowledgement of the discovery. This is because coordinating acknowledgement messages among multiple potential receivers can be  costly, requiring additional listening by the transmitter and possibly collision resolution.


\subsection{Analysis}
While \name can operate in general scenarios, for analytical tractability, we assume the following: 
\\ 
\noindent {\bf (A1)} All nodes are homogeneous, namely, have the same power budget $P_b$ and the same hardware.

\noindent {\bf (A2)} Every pair of nodes can exchange packets (clique
topology) with no packet errors due to noise.

\noindent {\bf (A3)} The number of nodes, $N$, is known a priori.

These assumptions are applicable to some systems and envisioned applications. For example,  when tracking boxes in a room (Fig.\ref{fig:intro}(a)), these assumptions are close to reality as nodes in close proximity  harvest similar amounts of energy, have few packets lost, and the number of nodes can be estimated a priori. However, for scenarios in which these assumptions do not hold, in Section~\ref{sect:dynamic}, we present \name-Dynamic which is based on relaxed assumptions and discuss the implications. \comment{Help with wording}

We note that, as this is the first attempt to develop an ND protocol explicitly for EH nodes, it is natural to consider the homogeneity assumption (A1). Additionally, several other works also make assumptions similar to (A2) and (A3) (e.g., \cite{Bakht_mobicom2012,sun2014hello} only consider ND for a link, $N=2$, with no collisions or packet loss).

Using these assumptions, we now use techniques from renewal theory~\cite{cinlar1969markov} to analyze \name~for a network of $N$ nodes. The renewal process is shown pictorially in Fig.~\ref{fig:netRen}. 
The renewal initiates with all nodes in the sleep state and ends after one node completes its transmission, whether the message is heard or not. The sleep duration for each node follows a memoryless exponential distribution. Therefore, for all analytical purposes, all nodes effectively initiate their sleep state at the start of the renewal.

In each renewal, the \emph{first} node to wake up begins its listen state, and after a duration $l$, it transmits its discovery message. This is exemplified by node 6 in Fig.~\ref{fig:netRen}; we denote by ${\cal N}_{t}$ the set containing a single transmitting node in a renewal.

Nodes that are in the receive state ($r$) when a message transmission begins, will
stay in this state until the transmission is completed and then switch to the sleep
state ($s$). We denote by ${\cal N}_{r}$ the set of such nodes and $|{\cal N}_{r}|$ the size of the set, exemplified by nodes 2-4 in Fig.~\ref{fig:netRen}.
The expected idle listening time of a node in ${\cal N}_{{r}}$ is
denoted by $\chi$. Fig.~\ref{fig:netRen} shows examples of idle listening durations for nodes 2--4, denoted as $\chi_i$. 
Any node which wakes up in the middle of the
message transmission immediately senses the busy channel
and returns to the sleep state. An example is node 5 in Fig.~\ref{fig:netRen}.

When the transmission is completed, all nodes are in sleep state and
the renewal restarts. The average renewal duration is the time it takes
for the first node to wake up (occuring with rate $ N\lambda$), listen for a duration $l$, and transmit a message for a duration of $M$. Hence, the
expected renewal duration $\rho$ is:
\begin{equation}
\rho = 1/(\lambda N) + l + M. \label{eqn:renewaldur}	
\end{equation}

\subsection{Discovery Rate}
Recall that the objective of \name~is to maximize the discovery rate, $U$ (see Def.\ \ref{defi:discrate}). Considering $U$
as the reward function and applying the elementary renewal theorem for
renewal-reward processes~\cite{cinlar1969markov}, we obtain:
\begin{equation}
U:= \lim_{t\rightarrow \infty} \frac{u(t)}{t} = \frac{\E[|{\cal N}_{{r}}|]}{\rho}, \label{eqn:objexp}
\end{equation} where $u(t)$ represents the number of discoveries (as defined
by Def.\ \ref{defi:disc}) by time $t$ and $\rho$ is computed by~\eqref{eqn:renewaldur}.

There are $N-1$ nodes who are not the transmitter in the renewal, each of which is equally and independently likely to discover the transmitter. 
A discovery occurs if the node wakes up from sleep within a period of time $l$ after the transmitting node (${\cal N}_{{t}}$) wakes up, an event with probability $1-\mathrm{e}^{-\lambda l}$. Hence,
\begin{equation}
 \quad \E[|{\cal N}_{{r}}|] =
(N-1)(1-\mathrm{e}^{-\lambda l}). \label{eqn:recvnum}
\end{equation}

\subsection{Energy Consumption}
\label{sect:energy}

Since all nodes are homogenous (A1), we let $n$ denote an arbitrary node and define a random variable $Y$ that indicates the set (${\cal N}_{{t}}$, ${\cal N}_{{r}}$) in which the node resides in the renewal:
\begin{equation}
Y = 0 \mbox{ if } n\in {\cal N}_{{t}}; Y=1 \mbox{ if } n\in {\cal N}_{{r}}; Y=2 \mbox{ else}.
\end{equation}

We let $\eta(y)$
represent the expected amount of energy (\si{\micro\joule}) consumed by a node in a renewal in which
$Y=y$. Thus,
\begin{equation}
\quad \eta(0) =  C_{sr}+ P_r l + P_t M  + C_{ts}\label{eqn:txener}, \\
\end{equation}
\begin{equation}
\quad \eta(1) = C_{sr}+P_r (\chi+M)  + C_{rs}\label{eqn:rxener}. \\
\end{equation}

Eq.~\eqref{eqn:txener} defines the energy consumption of the transmitting
node, which consumes energy to wake up from sleep ($C_{sr}$), listen for a period of $l$,
transmit a message of length $M$, and then return to sleep ($C_{ts}$). For a receiving node, the
expected energy consumption is defined in \eqref{eqn:rxener} and  consists
of idle listening before the message transmission (with $\chi$ denoting the expected duration of idle listening, shown in Fig.~\ref{fig:netRen}). Here, $\chi$ is a random variable denoting the duration during which a node idly listens to the channel before receiving a message. 
Without loss of generality, let us assume that the transmitter in a renewal (${\cal N}_{t}$, e.g., node $6$ from Fig.~\ref{fig:netRen}) enters the listen state at $t = 0$, and at $t = l$, it transmits the discovery message. Let $x$ denote the idle listening time for a given node where $x$ is exponentially distributed with $0 <x < l$.  We look to find $\chi = \E \left[x | x<l \right]$, i.e.,
\begin{align*}
\chi  = \int_{t=0}^{+ \infty} \textrm{Pr} (x>t | x<l)\, \textrm{d}t 
 = \int_{t=0}^{l} \frac{ (1 - e^{-\lambda l}) - (1 - e^{-\lambda t })}{1 - e^{-\lambda l}}\, \textrm{d}t = \frac{1}{\lambda} - \frac{l e^{-\lambda l}}{1 - e^{-\lambda l}}
\end{align*}

Then, the node listens for the duration of the message $M$. Throughout this paper, we assume that nodes which sleep for the entire renewal (e.g., nodes 1 and 7 in Fig.~\ref{fig:netRen}), and those which wake up briefly and sense a busy channel (e.g., node 5), do not consume power, and thus $\eta(2) = 0$. In Appendix~\ref{app:sleeping}, we show how it can be relaxed.

The computation of $\textrm{Pr}(Y=y)$ for $y=0,1$ is as follows. 
By definition of the renewal,
there will be exactly one transmitter in a renewal and due to assumption (A1), $\textrm{Pr}(Y=0)=1/N$. All the remaining $N-1$ nodes successfully receive the message, if they start listening in a period of length $l$ preceding the transmission. Hence, since the sleep duration is exponentially distributed, $\textrm{Pr}(Y=1)= (1-\mathrm{e}^{-\lambda l})
(N-1)/N$.

Define $\Phi(y) = \textrm{Pr}(Y=y) \eta(y)/\rho$ and note its units are (\si{\milli\watt});
we will often refer to $\Phi(0)$ as the {\em probing
power} while $\Phi(1)$ is referred to as
the {\em discovery power}. As described above, $\eta(2)=0$, and thus $\Phi(2)=0$. 
%
%
The expected power consumed in a renewal must meet the
power budget, 
$\Phi(0) + \Phi(1) \leq P_b$. 

\section{Optimization of~\name}
\label{sect:solutions}

Clearly, the choice of the sleep rate ($\lambda$) and the
listen duration ($l$) determines the power consumption of the node as well as the discovery rate $U$.  
First, we
demonstrate that an analytical solution
is difficult to obtain. Next, we
describe the \name Configuration Algorithm (PCA) which obtains the \emph{configuration parameters}
($\lambda$, $l$) for~\name. Finally, we demonstrate that
the PCA obtains a nearly-optimal discovery rate.


 \subsection{Problem Formulation and Preliminaries}

Finding $(\lambda^*,\ l^*)$ that maximizes $U$ is formulated as follows: 
\par\noindent
\begin{align}
\text{max}_{\lambda, l} & \quad U = (N-1) (1-\mathrm{e}^{-\lambda l})/ \rho  \label{eqn:obj_form}  \\
\text{s.t.}&  \quad    \Phi(0) + \Phi(1) \leq P_b \label{eqn:energy_form},
\end{align}
where \eqref{eqn:obj_form} is derived using \eqref{eqn:objexp} and \eqref{eqn:recvnum}.
Recall that $\rho$ is computed from~\eqref{eqn:renewaldur} and $\Phi(y)$ is computed using the results from Section~\ref{sect:energy}. The problem as formulated above is non-convex and non-linear, and is thereby challenging to solve.

In the following subsections, we will attempt to find nearly-optimal \name
configuration parameters ($\lambda, l$). 
We now provide several observations on the specific structure of the problem
which are used throughout this section. First, the following Taylor-series approximation is useful: 
\begin{equation}
\ \ \mathrm{e}^{-x} \geq 1-x \text{ for } x \geq 0, \text{ and }  \mathrm{e}^{-x} \approx 1-x \text{ for } x \approx{0} \text{.} 
\label{eqn:e_x_approx}
\end{equation}
We substitute $x$ with $\lambda l$ in~\eqref{eqn:e_x_approx},\footnote{Limited power budgets cause EH nodes to be in the sleep state much longer than in the listen state. Thus, $\lambda l \approx 0$ and \eqref{eqn:e_x_approx} is a good approximation.} 
\begin{align}
	U \leq  (N-1)\lambda l/\rho:=\overline{U}.\label{eqn:objupper}
\end{align}
\begin{wrapfigure}{}{3in}
\begin{minipage}{3in}
\vspace*{.3in}
\begin{algorithm}[H]
\footnotesize
\floatname{algorithm}{}
\renewcommand{\algorithmicrequire}{\textbf{}}
\renewcommand{\algorithmicensure}{\textbf{}}
\renewcommand{\thealgorithm}{}
\caption{\small \name Configuration Algorithm (PCA)}
\label{alg:EADCA}
\begin{algorithmic}[1]
\For{$K=[0,\epsilon, 2\epsilon,\ldots, \lfloor \frac {\rho_{\max}}
  {\epsilon} \rfloor \epsilon]$}
\State {Find $(\lambda,\ l)$ that maximize \eqref{eqn:objupper} s.t.~\eqref{eqn:energy_form_relaxed}}
\If {$\lambda$,$l$ satisfy \eqref{eqn:energy_form}}
\State {Compute the discovery rate $U$} \label{ln:U}
\EndIf
\EndFor
\State \Return {($\lambda$, $l$) that maximize $U$, denoted as $\lambda_A, l_A,$ and $U_A$.}
\end{algorithmic}
\end{algorithm}
\end{minipage}
\end{wrapfigure}







\subsection{\name~Configuration Algorithm (PCA)}
\label{sect:feasible}


The \name Configuration Algorithm (PCA) returns a 
configuration of $\lambda$ and $l$ that satisfy \eqref{eqn:energy_form}.
To find a configuration with the highest discovery rate, the PCA utilizes a relaxed problem formulation as follows. 
An upper bound on the discovery power, $\overline{\Phi}(1)$, is computed by using \eqref{eqn:e_x_approx} to obtain $(1-\mathrm{e}^{-\lambda l}) \leq \lambda l$, which leads to,
\begin{align}
\Phi(1) \leq \overline{\Phi}(1) := \frac{N-1}{N \rho} \lambda l \left( P_r (\chi+M) + C_{sr} + C_{rs} \right). \label{eqn:disc_energy_approx}	
\end{align}
The relaxed power budget constraint is then,
\begin{align}
\text{s.t.}&  \quad    \Phi(0) + \overline{\Phi}(1) \leq P_b \label{eqn:energy_form_relaxed}.
\end{align}

The PCA analytically computes the values of ($\lambda,l$) that maximize $\overline{U}$ by solving for $\lambda$ in terms of $l$ in \eqref{eqn:disc_energy_approx}, and then finding the critical points where $\textrm{d} \overline{U}/\textrm{d} l = 0$. For computation tractability, the PCA replaces $\chi$ with a constant $K$ in $\overline{\Phi}(1)$.
The PCA uses the fact that, in practice, a node's sleep time is upper bounded, introducing an upper bound on the renewal duration $\rho_{\max}$. Thereby, the PCA sweeps values between $0 \leq \chi \leq \rho_{\max}$, and returns the best solution (i.e., the one that maximizes $U$). We denote the discovery rate that the PCA obtains by $U_A$ and the configuration parameters 
by ($\lambda_A$, $l_A$).

\begin{figure*}
\centering
\subfigure[]{\label{fig:discenergy}\includegraphics[width = \figSize]{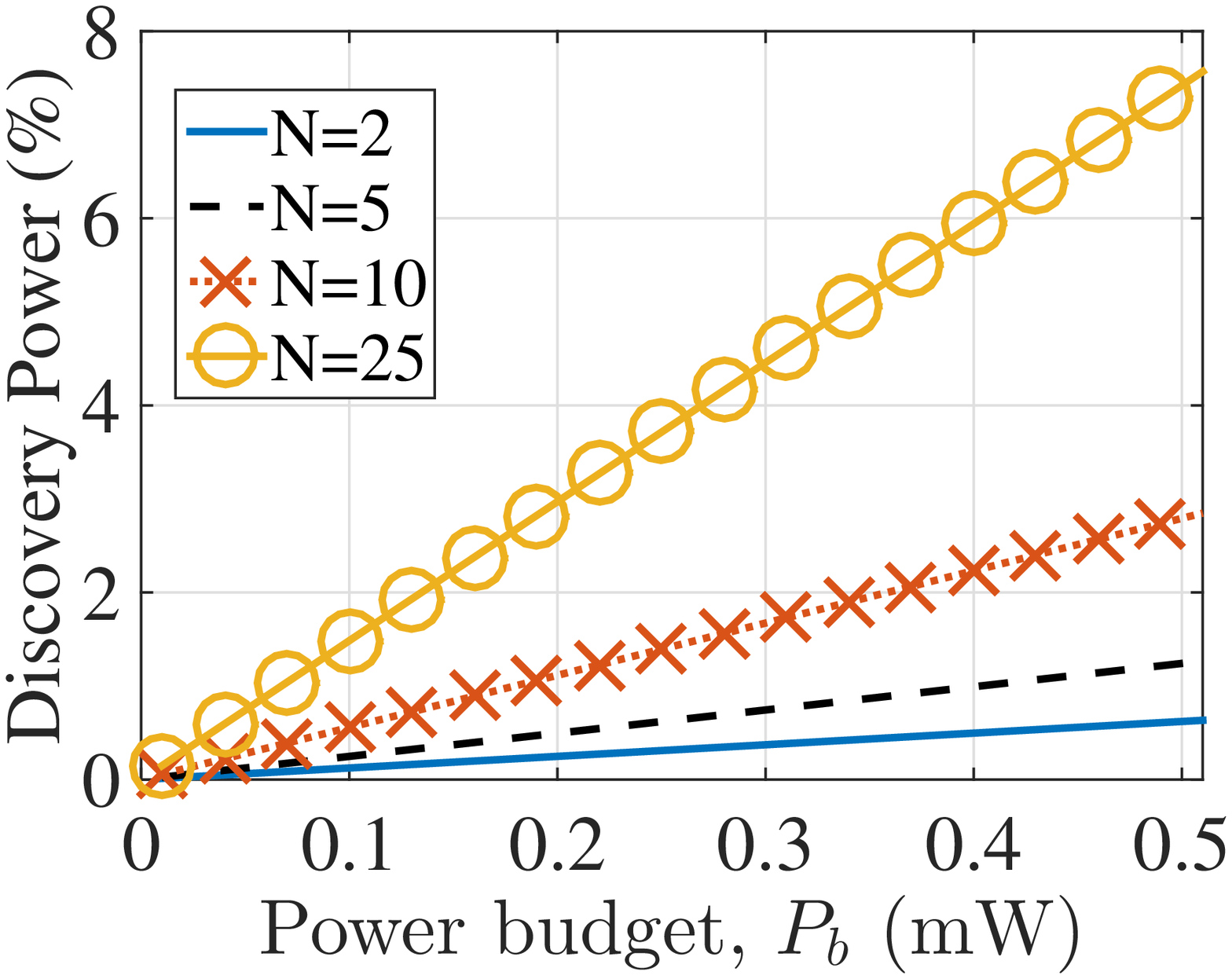}}
\subfigure[]{\label{fig:upperbound}\includegraphics[width =\figSize]{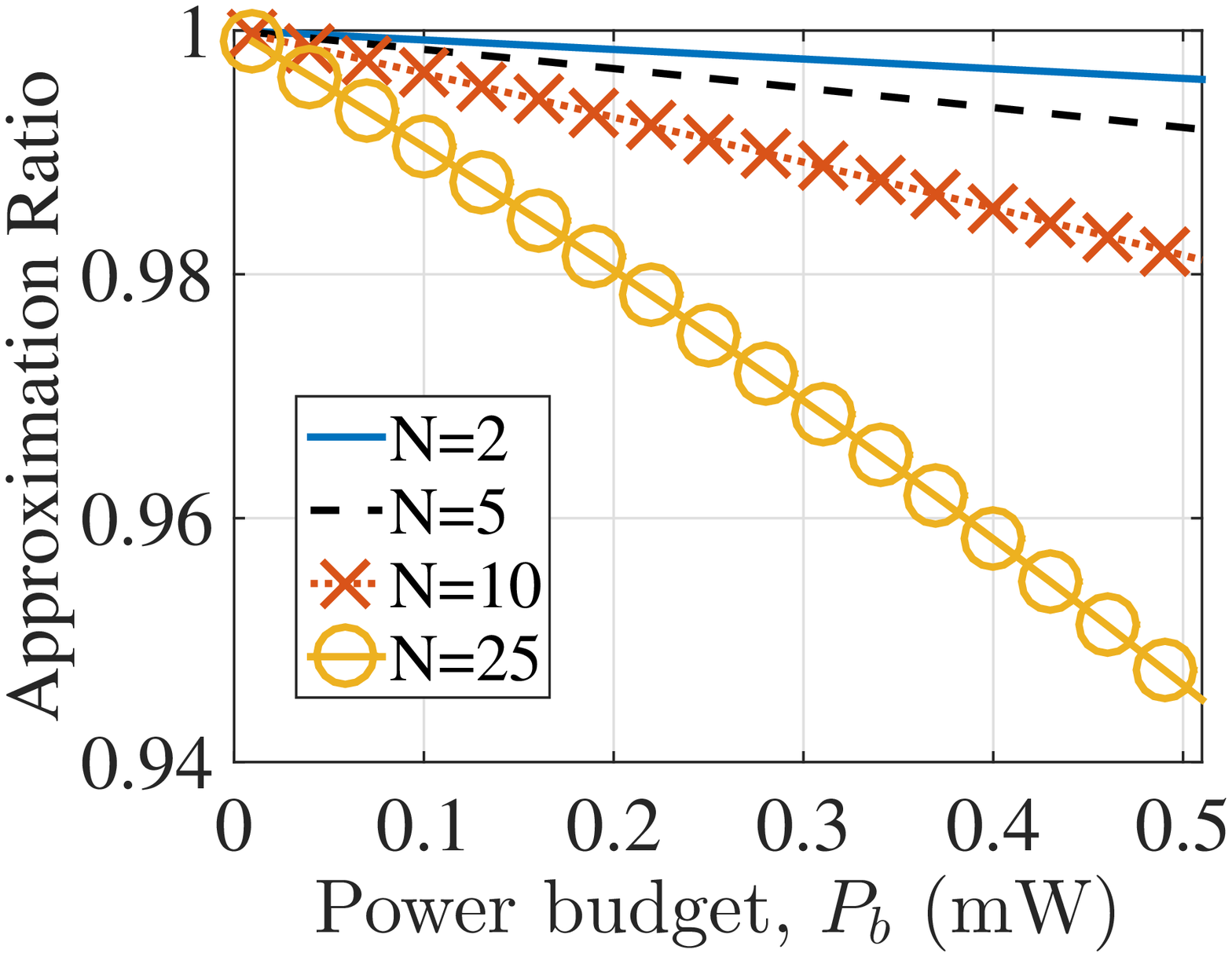}}
\subfigure[]{\label{fig:upper_opt_disc_rate}\includegraphics[width =\figSize]{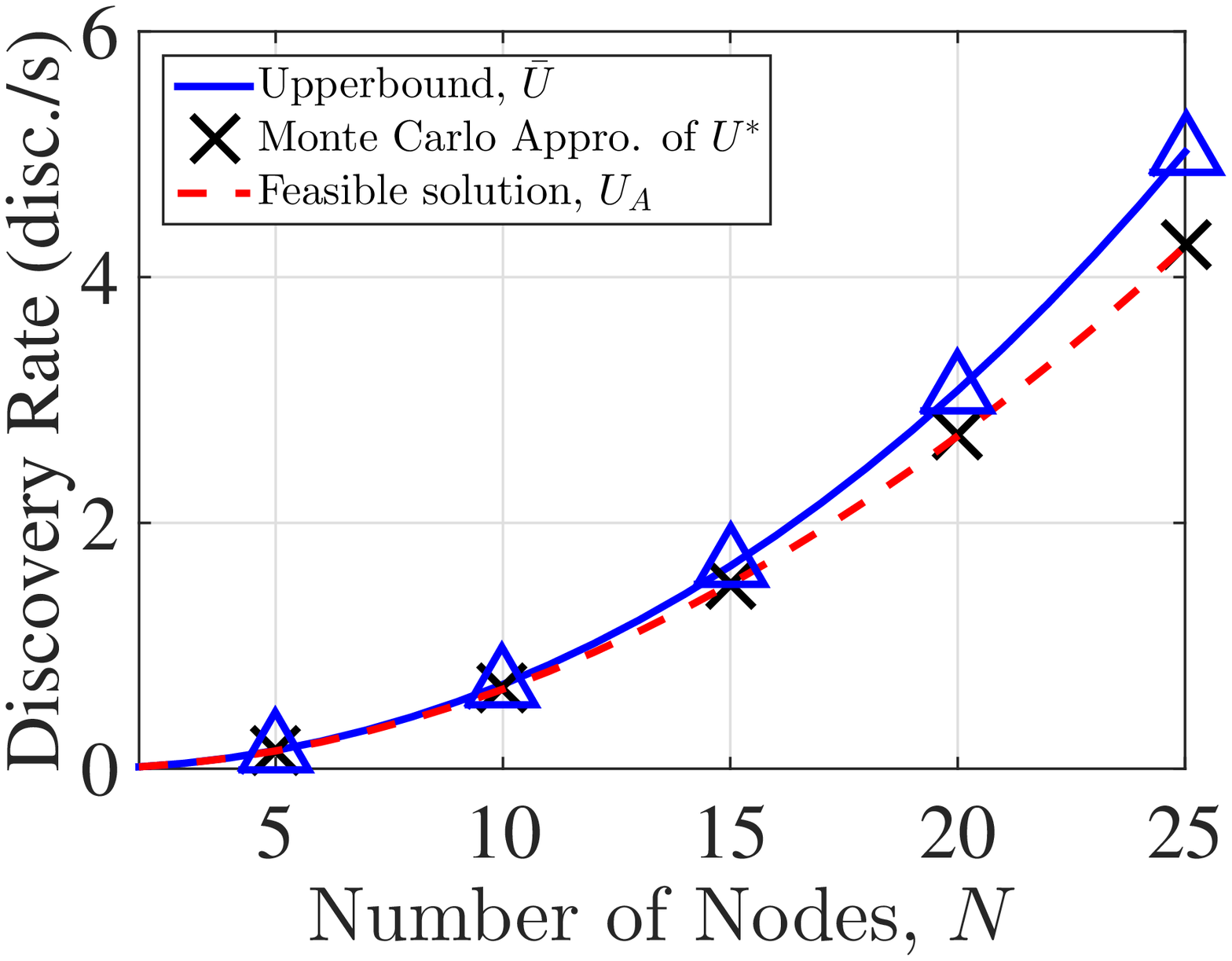}}
\vspace*{-.25in}
\caption{The performance of the PCA for varying power budgets, $P_b$, and number of nodes, $N$: (a) discovery power as the percentage of the power budget ($\Phi(1)$/$P_b$), (b) the approximation ratio (${U_A}/ {\overline{U^*}})$, and (c) the discovery rates resulting from the PCA ($U_A$) compared to the upperbound (${\overline{U^*}}$) and a Montee Carlo approximation of $U^*$.   \label{fig:perfupper}} 
\end{figure*}


\subsection{An Upper Bound}
\label{sect:upperbound}

To compute an {\em upper bound} on \name's optimal discovery rate, we first
derive a {\em lower bound} on the discovery power in an optimal solution, denoted by $\underline{\Phi^*}(1)$.

\begin{thm}
\label{thm:upper}
The discovery power in an optimal solution, $\Phi^*(1)$, satisfies,
\begin{equation}
  \Phi^*(1) \geq \frac{U_A}{N} (P_r M+ C_{rs}+C_{sr}) := \underline{\Phi^*}(1), \label{eqn:upper_disc_bound}
\end{equation}
where $U_A$ is the discovery rate returned by the PCA.
\end{thm}
\begin{IEEEproof}
The discovery energy for the optimal solution can be rewritten as
\begin{align*}
\Phi^*(1) = \underbrace{\frac{(N-1)(1-\mathrm{e}^{-\lambda^* l^*}) P_r \chi}{N\rho}}_{\geq 0} +
\underbrace{\frac{(N-1)(1-\mathrm{e}^{-\lambda^* l^*})}{\rho}}_{=U^*} \frac{P_r M + C_{rs}+C_{sr}}{N},
\end{align*}
where $l^*$ and $\lambda^*$ are an optimal configuration. By optimality, $U^* \geq U_A$ for any feasible configuration parameters leading to a discovery rate of $U_A$. Since, $\chi \geq 0$, the proof follows.
\end{IEEEproof}
%

Using \eqref{eqn:objupper} and \eqref{eqn:upper_disc_bound}, 
an upper bound optimization problem is formulated as, 
\par\noindent
\begin{align}
\max_{\lambda, l}: & \quad \overline{U^*} = (N-1) \lambda l/ \rho \label{eqn:objupper2} \\
\textrm{s.t.}: & \quad \Phi(0) + \underline{\Phi^*}(1) \leq P_b \label{eqn:energyCon'}.
\end{align}
Note that \eqref{eqn:energyCon'} effectively considers only a portion of
the discovery power.  This implies that the upper bound solution may be infeasible as it will incur an average power spending value higher than $P_b$. However, a solution maximizing \eqref{eqn:objupper2} is in fact an upper bound on the optimal solution $U^*$, and is denoted by $\overline{U^*}$.

To obtain an optimal solution for the upper bound $\overline{U^*}$, we
first solve for $\lambda$ with respect to the power constraint in (\ref{eqn:energyCon'}):
\begin{equation}
\lambda = P_b/\left(C_{rs} + C_{st} + P_{r} l + P_{t} M - N P_{b} \left(M + l\right)\right)
\label{eqn:tau_s_bound}
\end{equation}
Solving $\textrm{d} \overline{U^*}/\textrm{d} l = 0 $ obtains
the listen time to maximize $\overline{U^*}$,
\begin{equation}
l = \frac{\sqrt{(P_{r} \left(P_{r} - N P_{b}\right) \left(C_{rs} + C_{st} + P_{t} M\right) \left(C_{rs} + C_{st} + P_{t} M - M N P_{b}\right)}}{{P_{r}}^2 - N P_{r} P_{b}}
\end{equation}
Going forward, the corresponding value of the upper bound will be referred to as $\overline{U^*}$.
\comment{These equations need to include $U_A$}

\subsection{Performance of the PCA}
We now compare the discovery rate from the PCA ($U_A$) to the upper bound discovery rate computed in Section~\ref{sect:upperbound} ($\overline{U^*}$). In this section, we will refer to the ratio  $U_A/\overline{U^*}\leq 1$ as the \emph{approximation ratio}. 
Values close to 1 imply that $U_A$ is close to $\overline{U^*}$, and therefore, also close to the true optimal $U^*$.

Recall that the upper bound is computed by ignoring part of the discovery power, and therefore, violating the power constraint. Hence, when the discovery power $\Phi(1)$ in the optimal solution is indeed negligible ($\approx$ 0), the upper bound $\overline{U^*}$ is close to the true optimal, $U^*$. Therefore, as the discovery power decreases, the approximation ratio approaches 1.

In Fig.~\ref{fig:perfupper} we show both the discovery power and the approximation ratio resulting from the configuration parameters returned by the PCA for varying $N\in\{2,5,10,25\}$. In Fig.~\ref{fig:discenergy}, we consider the discovery power as a proportion of the total power budget $P_b$. As shown, smaller values of $N$ or $P_b$ result in a smaller proportion of discovery power.

Fig.~\ref{fig:upperbound} shows the approximation ratio $U_A/\overline{U^*}$ as a function of the power budget $P_b$. First, note that the approximation ratio is always greater than $94\%$ for all parameters considered. Since $U_A \leq U^* \leq \overline{U^*}$, the discovery rate provided by the PCA is within 6\% of the optimal. Additionally, for larger values of $N$ or $P_b$, the approximation ratio decreases. In this domain, the discovery power is larger (see Fig.~\ref{fig:discenergy}).
However, Fig.~\ref{fig:upper_opt_disc_rate} compares the discovery rate provided by the PCA to a Montee Carlo solution to \eqref{eqn:obj_form}.\footnote{The Montee Carlo solution generates over a $10^8$ random configuration parameters and returns the ($|lambda, l$) that satisfies \eqref{eqn:energy_form} with the largest discovery rate.} In all test cases considered, the PCA was within 0.25\% of the discovery rate of the Montee Carlo simulation. Therefore, although larger value of $N$ or $P_b$ result in PCA discovery rates ($U_A$) further from the upperbound ($\overline{U^*}$), in practice, the discovery rate from the PCA is very close to optimal ($U^*$). \comment{is this clear?}

\section{\name-Dynamic (\name-D)}
\label{sect:dynamic}
\name is analyzed assuming that nodes are homogenous (A1), are arranged in a clique (A2), and the number of nodes $N$ is known a priori (A3). However, when these assumptions do not hold, the expected power consumption of a node operating with \name (see Section~\ref{sect:energy}) will vary and the power budget is no longer satisfied. Therefore, in this section, we present \name-Dynamic (\name-D).


\name-D operates with the same behavior as \name, transitioning between the sleep, receive, and transmit states.
However, to handle the varying power consumption with the relaxed assumptions, the rate of the exponential sleep duration is dynamic, and is adapted based on the voltage of the capacitor,\footnote{A similar adaptation mechanism was also proposed in \cite{vigorito2007adaptive}.} $V_{\rm cap}$. Thereby, if a node consumes too much power, its voltage will decrease and it will adapt by staying in the sleep state for longer durations.


Formally, the configuration parameters for \name-D are computed as follows. In this case, $P_b$ represents an \emph{estimated} power budget for each node, yet we allow for each node to harvest power at varying rates around $P_b$.
The sleep duration is scaled such that the nodes' anticipated power consumption is 0.01\si{\milli\watt} when $V_{\rm cap}=3.6$\si{\volt}, and is $P_b$ when $V_{\rm cap}=3.8$\si{\volt}. 
From the two points, the desired power consumption of the node, $P_{\rm des}$,  is computed as a linear function of the capacitor voltage ($V_{\rm cap}$),
\begin{align*}
	P_{\rm des}(V_{\rm cap}) = \frac{P_b - 0.01}{3.8- 3.6}  \left( V_{\rm cap} - 3.6 \right) + 0.01,\, 3.6 \leq V_{\rm cap} \leq 4.
\end{align*}

Based on the desired power consumption $P_{\rm des}$, a node adjusts its sleep duration. As mentioned above, we cannot explicitly relate the sleep duration to the power consumption for a node. Instead, we estimate the power consumption by ignoring the \emph{discovery power}. That is, we assume that a node always follows the sleep, receive, transmit cycle and is spending on average at rate,
\begin{align*}
P_{\rm est} = \frac{\eta(0)}{1/\lambda + l + M} = \frac{P_r l + P_t M + C_{sr} + C_{ts}}{1/\lambda + l + M}.
\end{align*}
The average sleep duration, $1/\lambda$, is computed as a function of $V_{\rm cap}$ by solving $P_{\rm est} = P_{\rm des}$,
\begin{align}
1/\lambda = \frac{P_r l + P_t M + C_{sr} + C_{ts}}{P_{\rm des}(V_{\rm cap})} - l - M. \label{eq:blink-d}
\end{align}
We remark that the listen time $l$ is obtained using the PCA with $N = 2$ (i.e., we try to maximize the discovery rate for each {\it directional link}).

We claim that the robustness of \name-D is two-fold. First, it is power aware and nodes can operate under different and varying power harvesting rates, relaxing (A1). Additionally, it does not require any a priori knowledge of the size or topology of the network, relaxing (A2), (A3).



\section{Experimental Performance Evaluation}
\label{sect:experiments}
\begin{figure}[t]
\begin{minipage}{0.43\columnwidth}
\centering
\includegraphics[width=0.95\columnwidth]{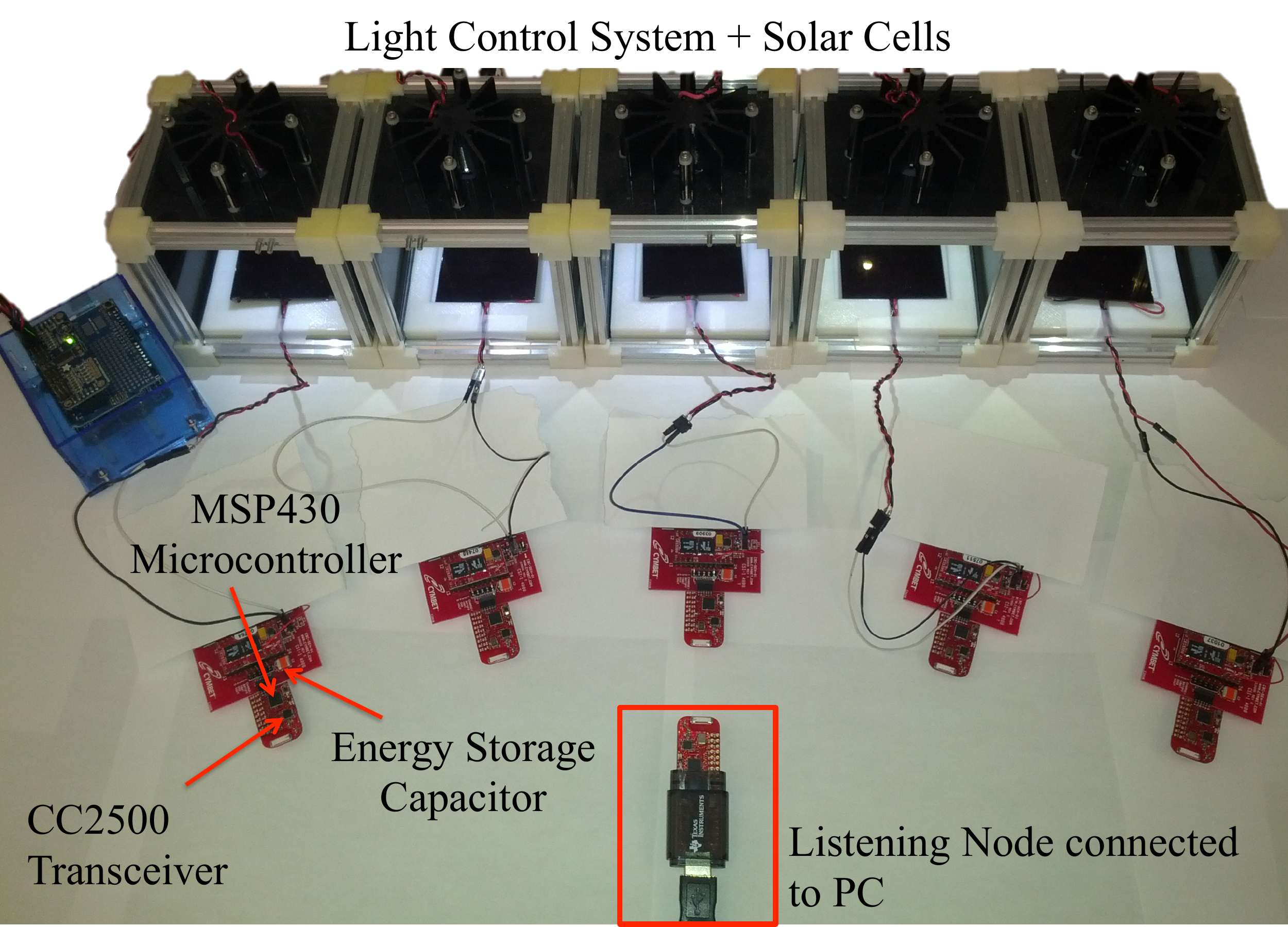}
\caption{\footnotesize \name experimental setup: 5 EH nodes harvest energy through the solar cells with neighbor discovery rates monitored by a listening node.  
\label{fig:NDsetup}}
\end{minipage}
\hspace*{0.005\columnwidth}
\begin{minipage}{0.28\columnwidth}
\centering
\scriptsize
\captionof{table}{Discovery message structure. \label{tab:discoverystruct}}
\vspace*{-.08in}
\begin{tabular}{ | c | l | }
\hline
Byte  & Data \\ \hline\hline
0     & Packet length (18 bytes) \\ \hline
1     & Type \\ \hline
2-11  & Neighbor table \\ \hline
12-13 & Capacitor voltage  \\ \hline
14-15 & Debugging information \\ \hline
16    & Transmissions counter \\ \hline
17    & Originating node ID \\
\hline
\end{tabular}
\vspace*{0.7in}
\normalsize
\end{minipage}
\hspace*{0.01\columnwidth}
\begin{minipage}{0.2\columnwidth}
\scriptsize
\centering
\captionof{table}{Measured\\ prototype parameters. \label{tab:enercosts}}
\vspace*{-.05in}
\begin{tabular}{| c | l |}
\hline
Param. & Value \\ \hline \hline
$P_{t}$ & 59.23\si{\milli\watt}\\ \hline
$P_{r}$ & 64.85\si{\milli\watt}\\ \hline
$M$ &  0.92\si{\milli\second}\\ \hline
$C_{sr}$ &  74.36\si{\micro\joule} \\ \hline
$C_{rs}$ &  13.48\si{\micro\joule}\\ \hline
$C_{tr}$ & 4.83\si{\micro\joule}\\ \hline
\end{tabular}
\vspace*{0.15in}
\vspace*{0.7in}
\normalsize
\end{minipage}
\end{figure}

We now evaluate \name using a testbed, pictured in Fig.~\ref{fig:NDsetup}, composed of TI eZ430-RF2500-SEH \cite{TIehavkit} prototypes (described in Section~\ref{sec:prototype}). 
First, we evaluate \name in the context of the model presented in Section~\ref{sect:sys_model}. We compare \name's experimental discovery rate, denoted by $U_E$, to related work.
Additionally, we present \name's performance with varying parameters (e.g., transmission power, message length).
 Then, we evaluate \name-D in scenarios with non-homogeneous power harvesting and multihop topologies. 

\begin{figure}[t]
\centering
\subfigure[\label{fig:txrxchar2}]{\includegraphics[width =  \figSize]{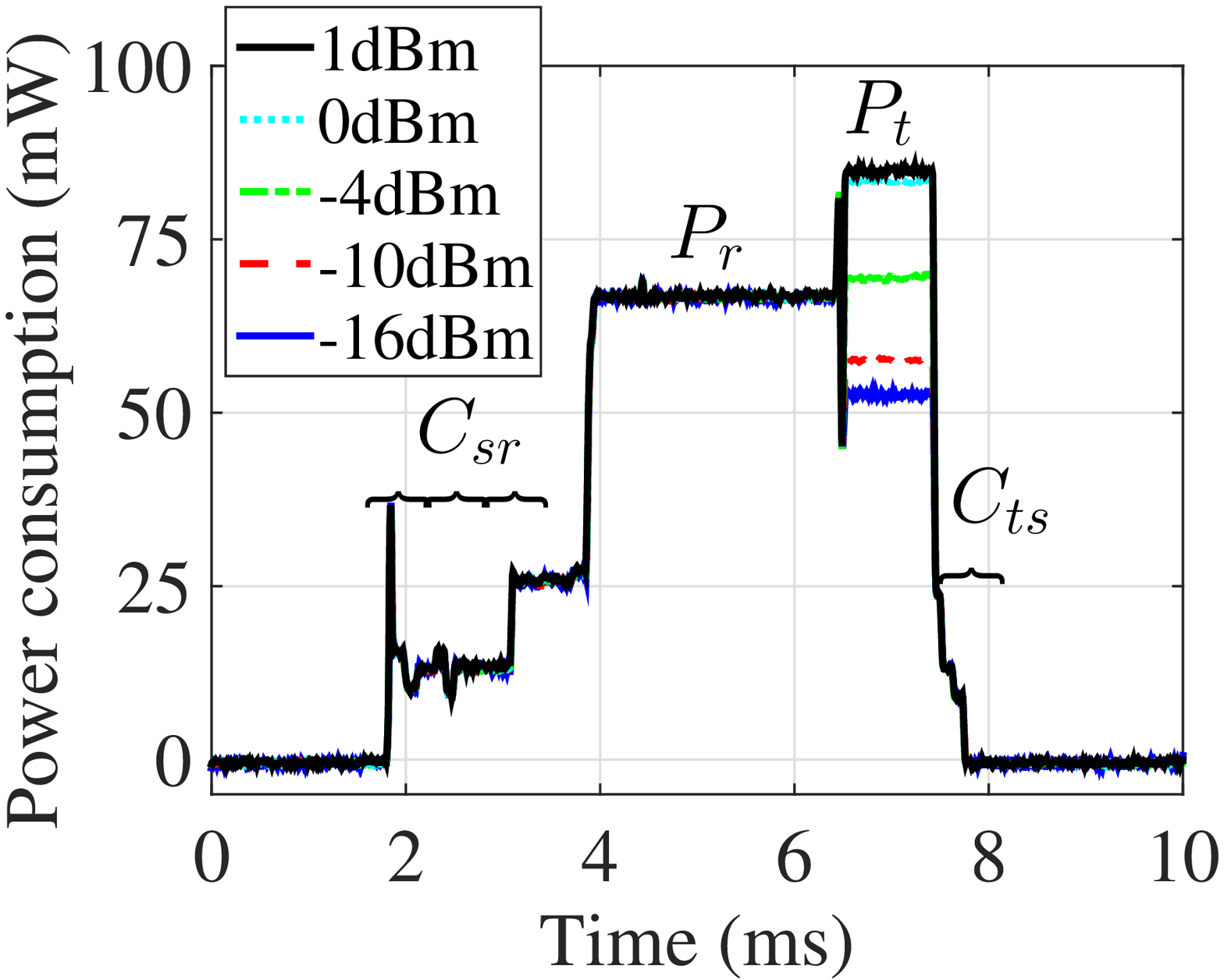}}
\subfigure[\label{fig:RateConverg}]{\includegraphics[width= \figSize]{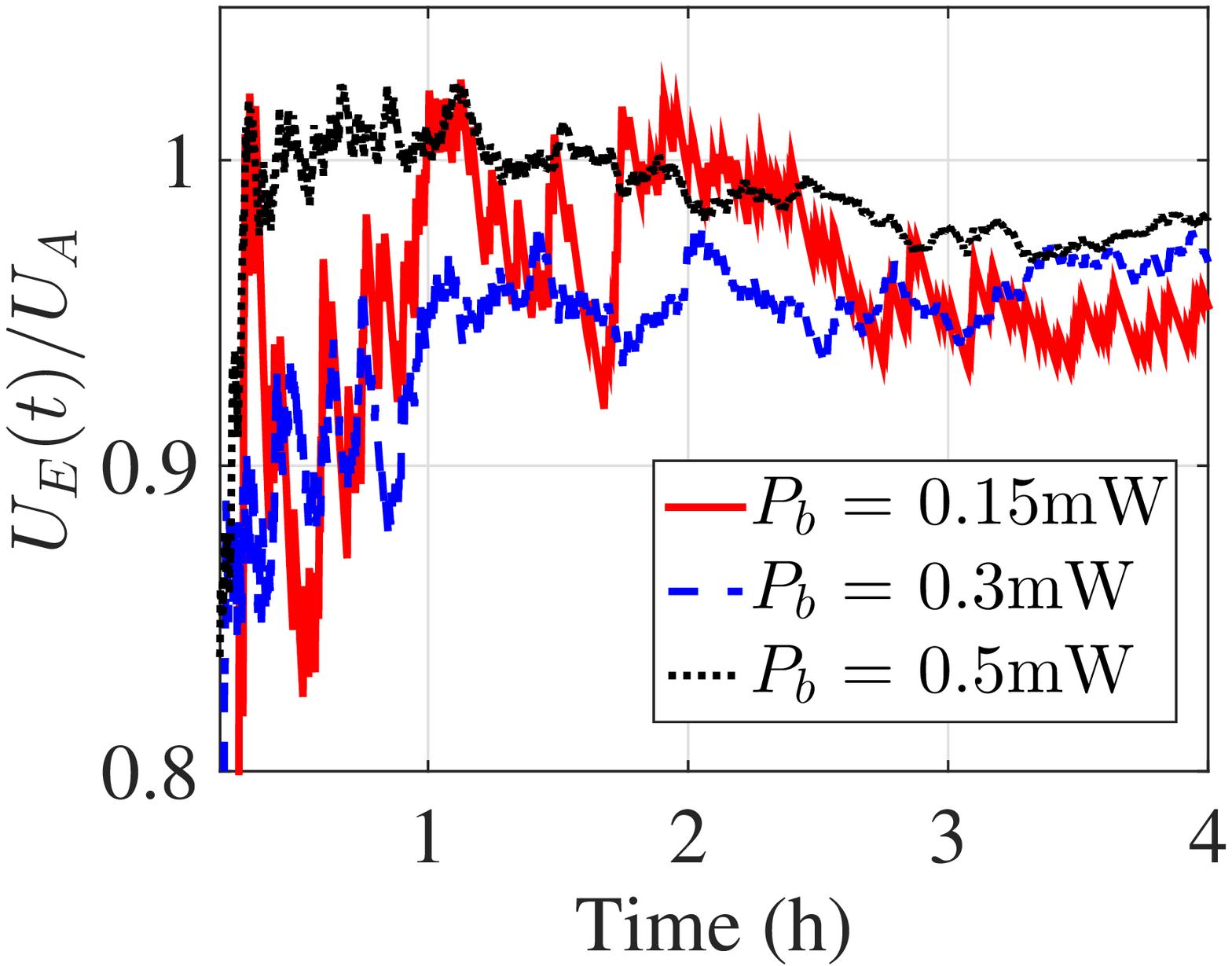}}
\subfigure[\label{fig:DiscRate}]{\includegraphics[width= \figSize]{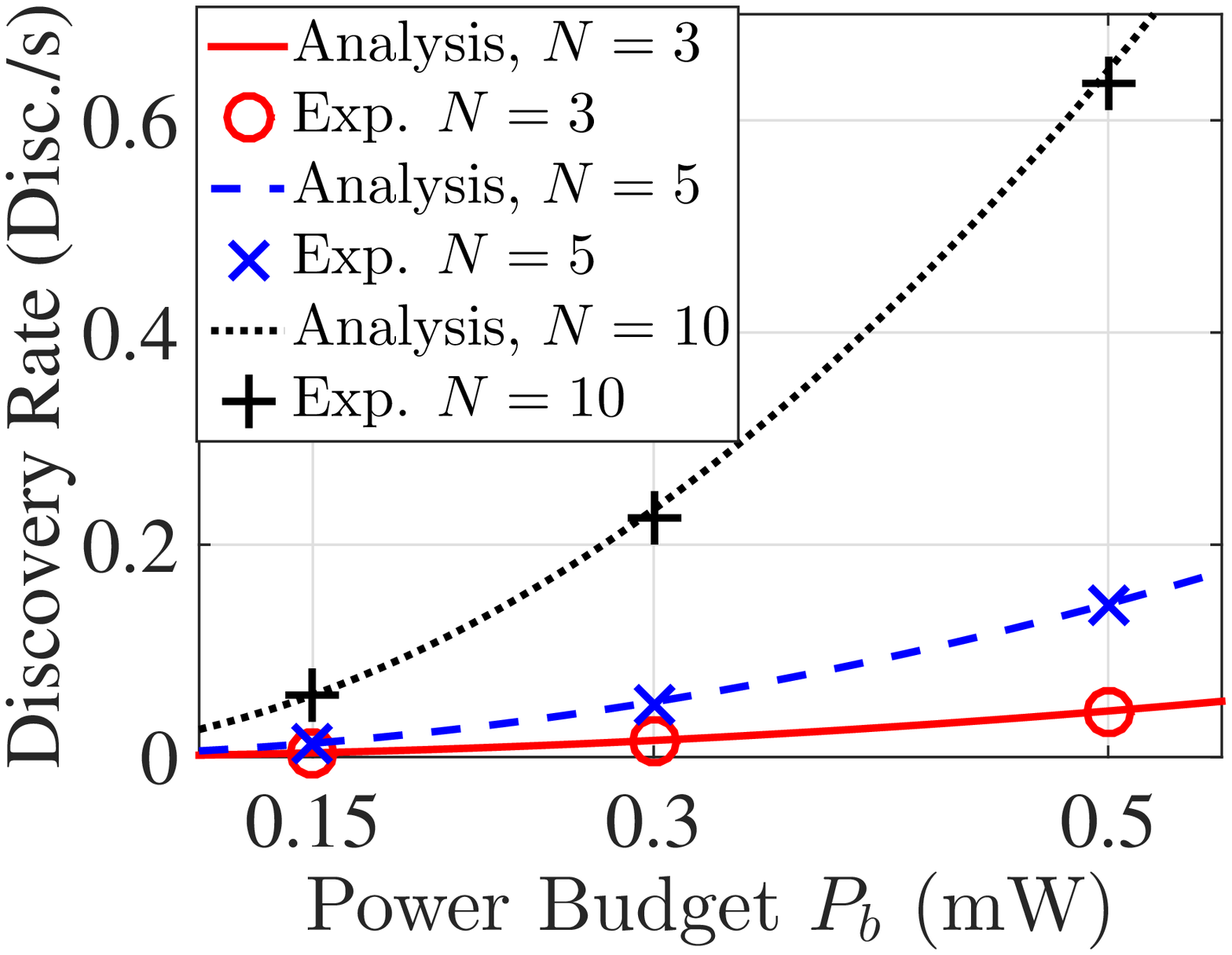}}
\vspace*{-.2in}
\caption{\footnotesize (a) Power consumption and transition costs for different transmission power levels for a node transitioning between the sleep, receive, and transmit states, and back to sleep. (b) \name experimental evaluation with varying power budgets, $P_b$: Convergence of the experimental discovery rate ($U_E$) to the analytical discovery rate ($U_A$) for $N=5$. (c) \name's discovery rate with varying power budgets, $P_b$, and number of nodes, $N$}
\end{figure}

\subsection{Protocol Implementation}
In accordance with \name, the microcontroller builds the discovery message and sends it to the low-power transceiver.
Table \ref{tab:discoverystruct} illustrates the structure of the discovery message.
The message contains debugging information, the source ID of the transmitting node,
and the node's capacitor voltage (which is sampled from the ADC).
Additionally, the 
message includes the number of discoveries from each neighbor since the initialization of the experiment, referred to as the node's {\em neighbor table}.
The total length of a discovery message is 18 bytes and the resulting transmission duration of the discovery message is $0.92$\si{\milli\second}. 

In order to characterize the energy costs, we measure the power consumption of the microcontroller and transceiver using an oscilloscope. 
Fig.~\ref{fig:txrxchar2} shows the power levels for a node transitioning between the sleep, receive, and transmit states. We compute the average power consumption and transition energy for each state, with values 
summarized in Table~\ref{tab:enercosts}.

We note that the transition times to and from the sleep state are non-negligible (in some cases a few ms). To account for this, these transition times are considered as part of the sleep state and, are therefore, subtracted from the actual sleep duration. We elaborate further on the importance of incorporating these switching costs in
Appendix~\ref{app:switching}.

The parameters in Table~\ref{tab:enercosts} compose the inputs to the PCA, which computes the rate of the exponential sleep $\lambda_A$ and the duration of the listen state $l_A$ as well as an expected discovery rate $U_A$. These \emph{configuration parameters} are loaded into the nodes for experimental evaluation in which we observe the discovery rate as well as the power consumption.

\subsection{Testbed and Experimental Setup}
\label{sect:testbed}

We consider networks of 3, 5, and 10 nodes ($N=3,5,10$). We consider power budgets of $P_b=0.15$, $0.3$, $0.5$mW; these are aligned with other solar harvesting budgets \cite{Gorlatova_NetworkingLowPower}. Initially, to confirm the practicality of \name when assumptions (A1), (A2), and (A3) hold, we place the nodes in close proximity with a homogenous power budget. In Section~\ref{sect:practical}, we will evaluate \name-Dynamic (\name-D) and relax these assumptions by considering a multihop topology and non-homogenous power harvesting. 

To facilitate experimental evaluation with up to $N=10$ nodes, 
in addition to an EH node shown in Fig.~\ref{fig:EHBprototype}(b), we also incorporate nodes powered by AAA batteries into the experiments. Both the EH node and the nodes powered by AAA batteries operate using the same configuration parameters and have identical behaviors (i.e., the source of power does not affect the behavior of \name). However, we carefully logged the power consumption of the EH node by including control information in the discovery message~(see Table~\ref{tab:discoverystruct}).

We utilize a listening node consisting of a microcontroller and transceiver set to a promiscuous sniffing mode to log experimental results. 
Powered by a USB port on a monitoring PC, the listening node reports all received messages to the PC for storage and post processing.
The experimental discovery rate, $U_E$, is computed by dividing the total number of discoveries since the initialization of the experiment by the experiment duration.
Clearly, the time until which the experimental discovery rate converges depends on the rate of discovery.
In Fig.~\ref{fig:RateConverg}, we observe the experimental discovery rate, $U_E$ over time for $N=5$ and $P_b=0.15, 0.3, 0.5$\si{\milli\watt}.  Based on the results shown in Fig.~\ref{fig:RateConverg}, all experiments were conducted for up to 96 hours. 

The light levels are set to correspond to each of the power budgets, $P_b$. However, the performance of the solar cells vary significantly due to external effects such as aging, orientation, and temperature~\cite{Margolies_EnHANTS_TOSN}. To mitigate these affects and facilitate repeatable and controllable experiments, we designed a software controlled light system which we describe in Appendix~\ref{app:lights}.

Additionally, as mentioned in Section~\ref{sect:model}, the prototype is not power aware. That is, although we can accurately measure the power harvested by the solar cell, it is difficult to control the energy actually stored in the capacitor, due to numerous inefficiencies of the harvesting circuitry, which are further described in Appendix~\ref{app:lights}.
As such, we empirically estimated the harvesting inefficiency to be 50\% and adjust the light levels to provide each node energy according to the value of $P_b$ chosen.
%

\begin{figure}[t]
\begin{minipage}{0.6\columnwidth}
\centering
\scriptsize
\captionof{table}{\footnotesize \name experimental parameters: $(\lambda_A, l_A)$ generated using the PCA 
for every input ($N, P_b$) pair and the resulting analytical ($U_A$) and experimental ($U_E$) discovery rate.\label{tab:solutions} }
\vspace*{-.05in}
\resizebox{\columnwidth}{!}{
\begin{tabular}{| c | c | c | c | c | c | c | c | c |}
\hline
N & \parbox{1.2cm}{\centering$P_b$ (mW)} & $\lambda_A^{-1}$(ms) & $l_A$(ms) & \parbox{0.7cm}{\vspace*{0.05cm}\centering Duty Cycle (\%)\vspace*{0.05cm}} & \parbox{0.75cm}{\centering $U_A$ (Disc./s)} & \parbox{0.8cm}{\centering $U_E$ (Disc./s)} & \parbox{0.6cm}{\centering Error (\%)} & \parbox{0.4cm}{\centering Run Time (h)}\\ \hline
\multirow{3}{*}{\centering 3} &	0.15 & 1778.68 & 2.066 & 0.168 & .0039 & .0038 & -1.35 & 36 \\  \cline{2-9}
                              & 0.3  & 887.39  & 2.070 & 0.336 & .0156 & .0154 & -1.23 & 36 \\ \cline{2-9}
                              & 0.5  & 530.88  & 2.075 & 0.561 & .0434 & .0438 & 1.07  & 48 \\ \hline
\multirow{3}{*}{\centering 5} &	0.15&1777.18&2.068&0.168 &.0130&.0132&1.43&96\\   \cline{2-9}
	& 0.3&885.91&2.075&0.337 &.0519&.0518&-0.33&60\\ \cline{2-9}
	& 0.5&529.43&2.084&0.564 &.1443&.1427&-1.15&18\\ \hline
\multirow{3}{*}{\centering 10} & 0.15&1773.49&2.075&0.169 &.0584&.0589&0.89&18\\  \cline{2-9}
	& 0.3&882.32&2.089&0.340 &.2332&.2341&0.38&18\\  \cline{2-9}
	& 0.5&525.97&2.107&0.572 &.6470&.6510&0.62&18\\ \hline
\end{tabular}}
\normalsize
\end{minipage}
\hfill
\begin{minipage}{0.35\columnwidth}
\centering
\scriptsize
\captionof{table}{\footnotesize Neighbor table for $N=5$, $P_{b} = 0.3$\si{\milli\watt} after 4 hours. Entry ($i$,$j$) shows the number of discoveries of node $j$ by node $i$. 
\label{tab:neighbortable}}
\vspace*{-.05in}
\resizebox{\columnwidth}{!}{
\begin{tabular}{| c | c | c | c | c | c | c |}
\hline
\parbox{0.4in}{\bf \vspace*{0.03in} \centering ND Table\vspace*{0.03in}}& {\bf 1} & {\bf 2} & {\bf 3} & {\bf 4} & {\bf 5} & {\bf Total RX} \\ \hline
{\bf 1} & 0  & 35 & 42 & 32 & 42 & 152 \\ \hline
{\bf 2} & 24 & 0  & 23 & 45 & 38 & 130 \\ \hline
{\bf 3} & 39 & 36 & 0  & 21 & 33 & 129 \\ \hline
{\bf 4} & 36 & 35 & 46 & 0  & 32 & 149 \\ \hline
{\bf 5} & 38 & 42 & 42 & 42 & 0  & 164 \\ \hline
\end{tabular}}
\vspace*{0.8in}
\normalsize
\end{minipage}
\end{figure}


\subsection{Discovery Rate}
For each ($N,P_b$) pair, we evaluate \name, with the experimental parameters summarized in Table \ref{tab:solutions}. 
  First, we note that \name's duty cycle is typically between 0.1--0.6\%, which is significantly lower than the duty cycles considered in related protocols~\cite{Sun_NDSurvey}. Additionally, note the accuracy of the analytical discovery rate, $U_A$, computed from~\eqref{eqn:objexp}, compared to the experimental discovery rate, $U_E$. On average, the error between them is $\approx 1\%$. This confirms the practicality of \name and the 
  model described in Section~\ref{sec:node_model}. 
  \comment{Gil wants more clarification here}

In Fig.~\ref{fig:DiscRate}, we plot the experimental and analytical discovery rate for each value of ($N,P_b$) shown in Table~\ref{tab:solutions} and observe the effect of varying $N$ and $P_b$. As expected, the discovery rate increases as $P_b$ increases. 
The number of nodes $N$ is directly correlated with the discovery rate, as indicated in~\eqref{eqn:objexp} and \eqref{eqn:recvnum}. 
As such, the discovery rate increases as $N$ increases.
%
%
%
%
%
%
%

Additionally, by tracking each nodes' neighbor table in Table~\ref{tab:neighbortable}, 
we confirm that all nodes discover one another and exhibit similar per link discovery rates.

\subsection{Discovery Latency and Comparison to Related Work}
\begin{figure*}[t]
\centering
\subfigure[\label{fig:discLatency}]{\includegraphics[width=\figSize]{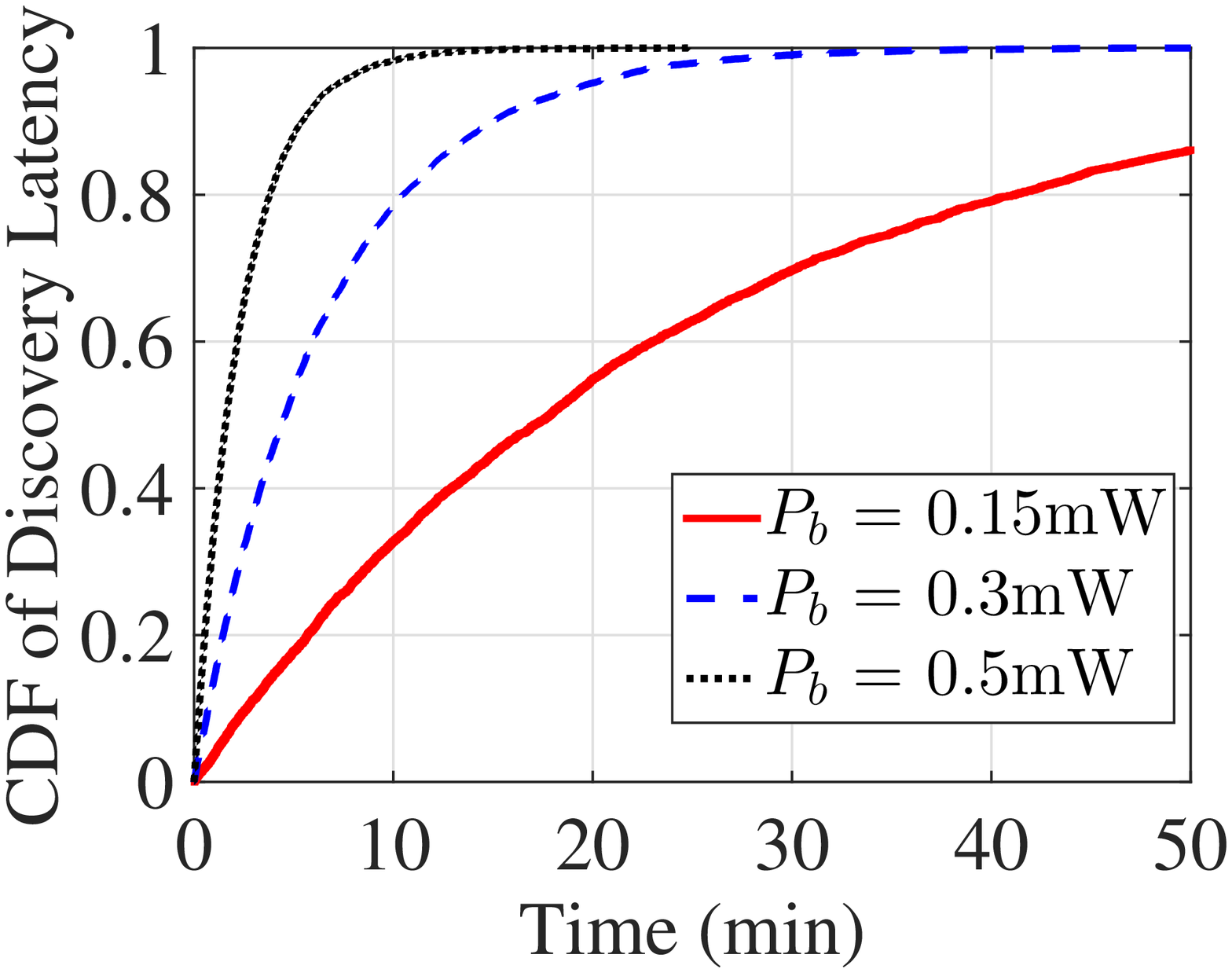}}
\subfigure[\label{fig:relatedDiscRate}]{\includegraphics[width=\figSize]{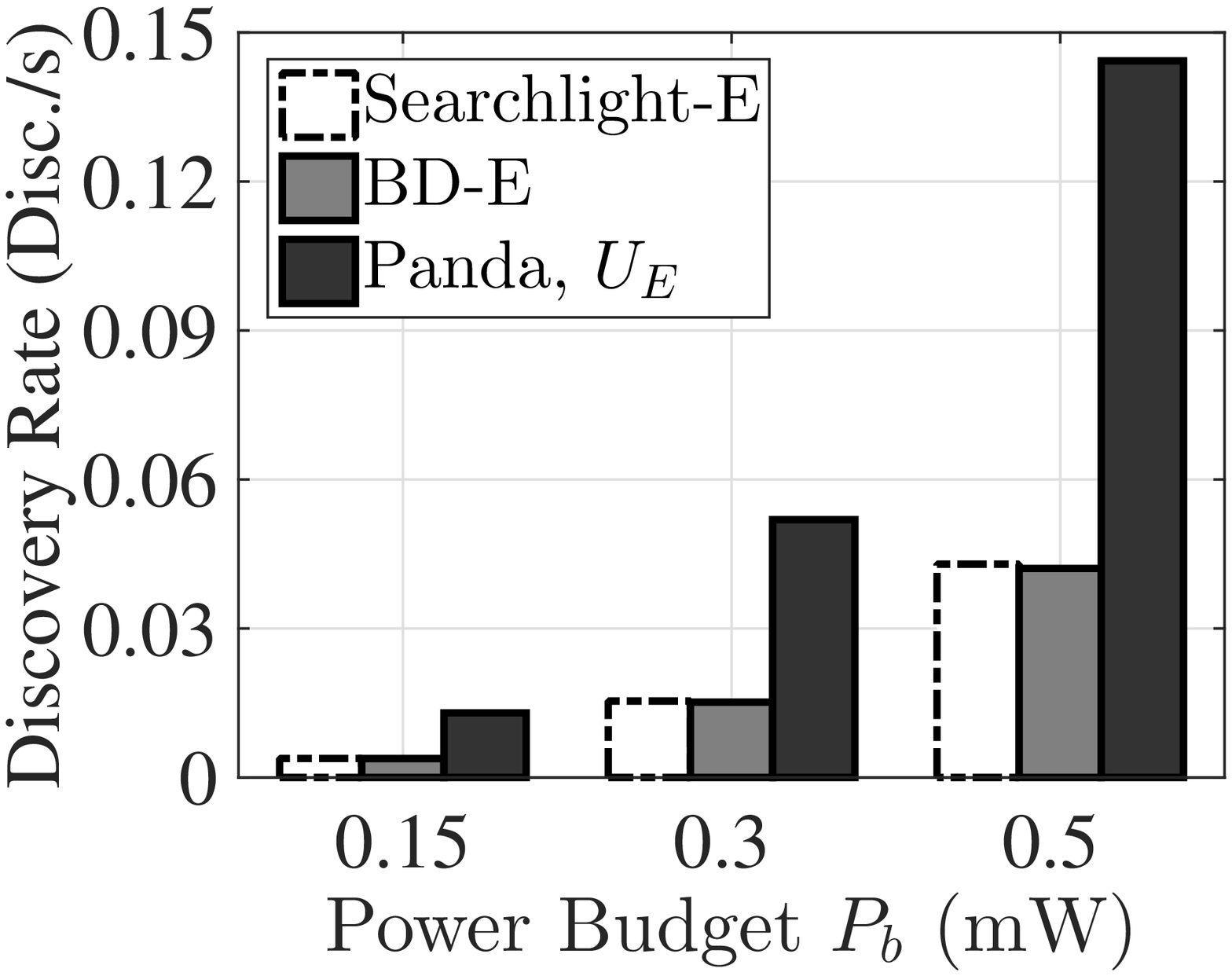}}
\subfigure[\label{fig:relatedMaxLatency}]{\includegraphics[width=\figSize]{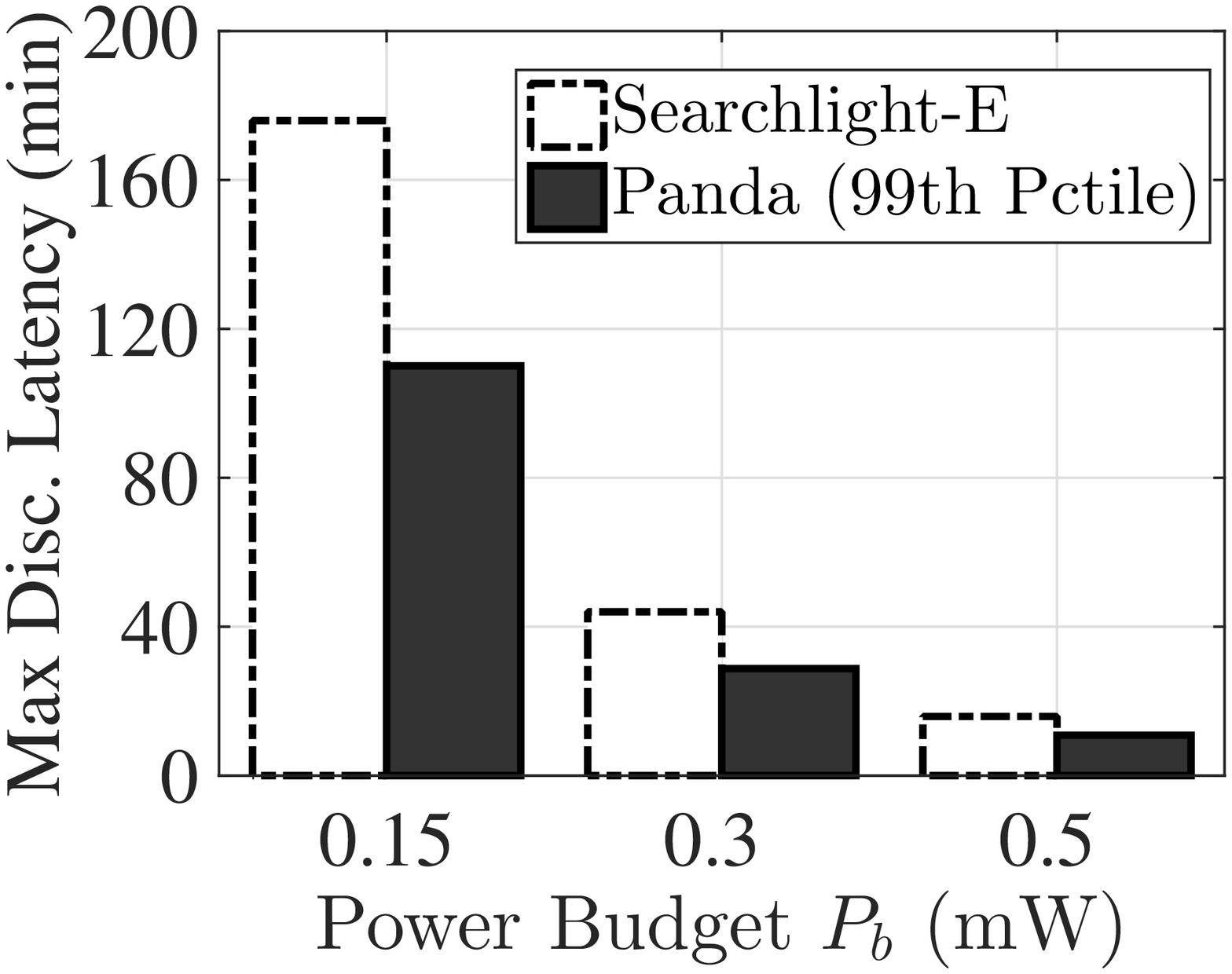}}	
\vspace*{-.2in}
\caption{\name experimental evaluation for $N=5$: (a) CDF of per link discovery latency; comparison to SearchLight-E~\cite{Bakht_mobicom2012} and BD-E~\cite{McGlynn_mobihoc01} of (b) the discovery rate and (c) the worst case latency.}
\end{figure*}

The discovery latency is the time between consecutive discoveries for a \emph{directional link}. It can be an important parameter for numerous applications where nodes are only within communication range for short periods of time.  Although the objective of \name is to maximize the discovery rate, in Fig.~\ref{fig:discLatency}, we show the CDF of the discovery latency for each \emph{directional link} in an experiment with $N=5$ and varying power budgets. Clearly, the average discovery latency decreases as the average discovery rate increases. Thus, for a higher power budget, the discovery latency decreases.

Previous work~\cite{Bakht_mobicom2012,sun2014hello,Dutta_disco08} focused on minimizing the worst case discovery latency for a link. We compare the discovery latency of \name, shown in Fig.~\ref{fig:discLatency}, to previous work. However, as mentioned in Section~\ref{sect:related}, previous work considers a duty cycle constraint instead of a power budget ($P_b$). To provide a means of comparison, we use the following equation to relate the power constraint to a duty cycle. 
\begin{align}
P_b 
&= 	\mbox{Duty Cycle(\%)} \cdot \mbox{Average Active Power (\si{\milli\watt})} \label{eqn:related}
\end{align}
We compare to the deterministic Searchlight protocol~\cite{Bakht_mobicom2012}, which  minimizes the worst case discovery latency\cite{sun2014hello}. We also compare to the the well-known probabilistic Birthday (BD) protocol~\cite{McGlynn_mobihoc01}.
To account for the power budget, we modify these protocols based on \eqref{eqn:related} (with details explained in Appendix~\ref{app:related})
 and denote them as Searchlight-E and BD-E. Based on previous work~\cite{sun2014hello}, we set the slot size for Searchlight-E and BD-E to 50\si{\milli\second} and add an overflow guard time of 1\si{\milli\second}.

In Fig.\ \ref{fig:relatedDiscRate}, we compare the average discovery rate for \name vs.\ simulations of the Searchlight-E and BD-E protocols. We found that \name typically outperforms the Searchlight-E and BD-E protocols by over 3x in terms of the average discovery rate.\footnote{As described in Appendix~\ref{app:related}, the simulations of Searchlight-E and BD-E do not account for packet errors or collisions. As such, the discovery rates for these protocols is likely to be lower in practice.}

Furthermore, in Fig.\ \ref{fig:relatedMaxLatency}, we consider the \emph{worst case} discovery latency and show that although \name has a non-zero probability of having any discovery latency, for the experiments we considered, the $99^{\rm th}$ percentile of discovery latency outperformed the Searchlight-E protocol worst case bound by up to 40\%.

Note that the Searchlight protocol was proven to minimize the worst case discovery latency. However, as shown through our evaluation, \name outperforms Searchlight-E by a factor of 3x in terms of average discovery rate. Moreover, in most cases (over 99\%), the discovery latency is below the worst case bound from Searchlight-E. This emphasizes the importance of incorporating a detailed power budget, as is done in \name, as opposed to a duty cycle constraint.

\subsection{Power Consumption}
\begin{figure}[t]
\centering
\subfigure[\label{fig:voltage}]{\includegraphics[width=\figSize]{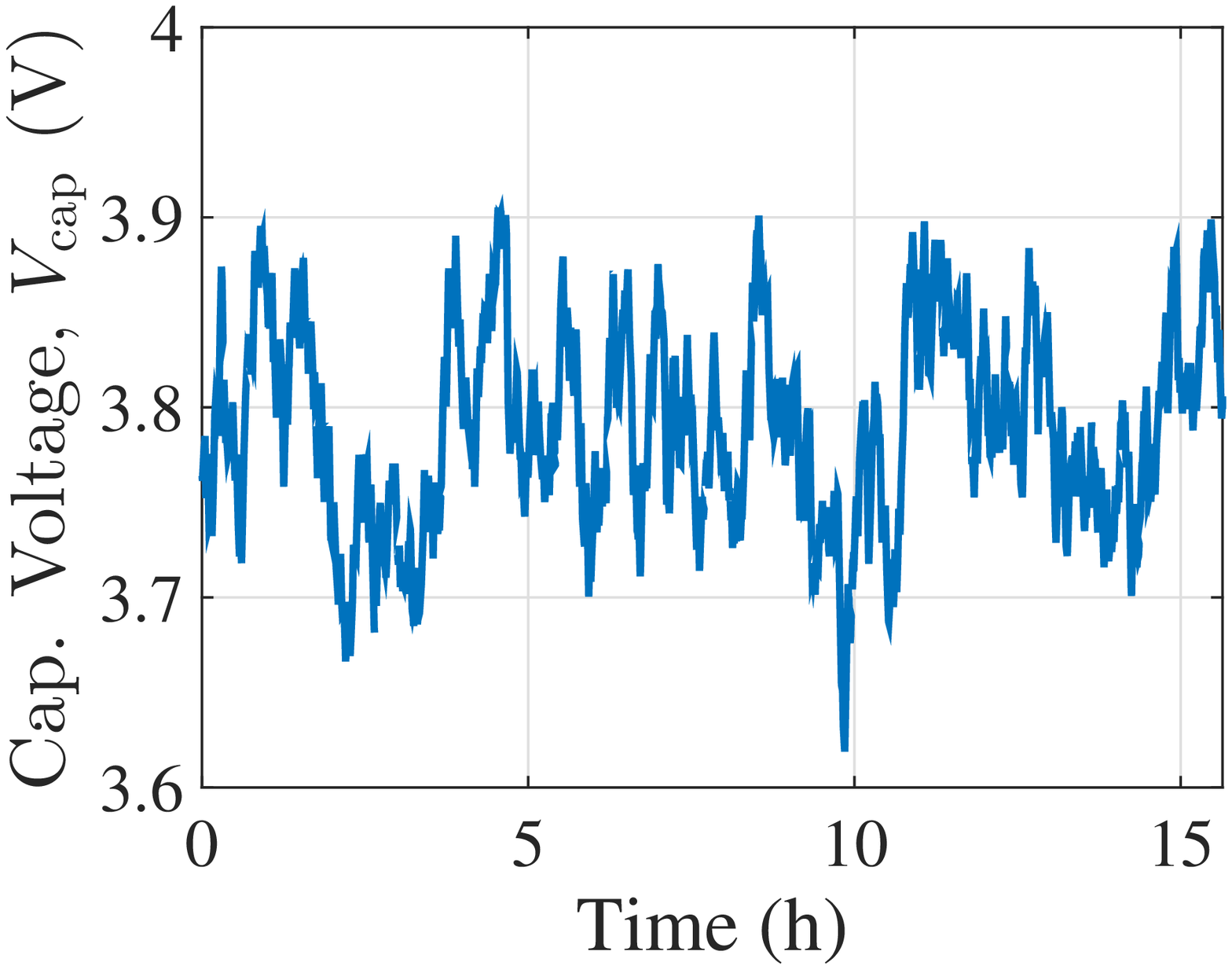}}
\hspace*{0.05\columnwidth}
\subfigure[\label{fig:varyCap}]{\includegraphics[width=\figSize]{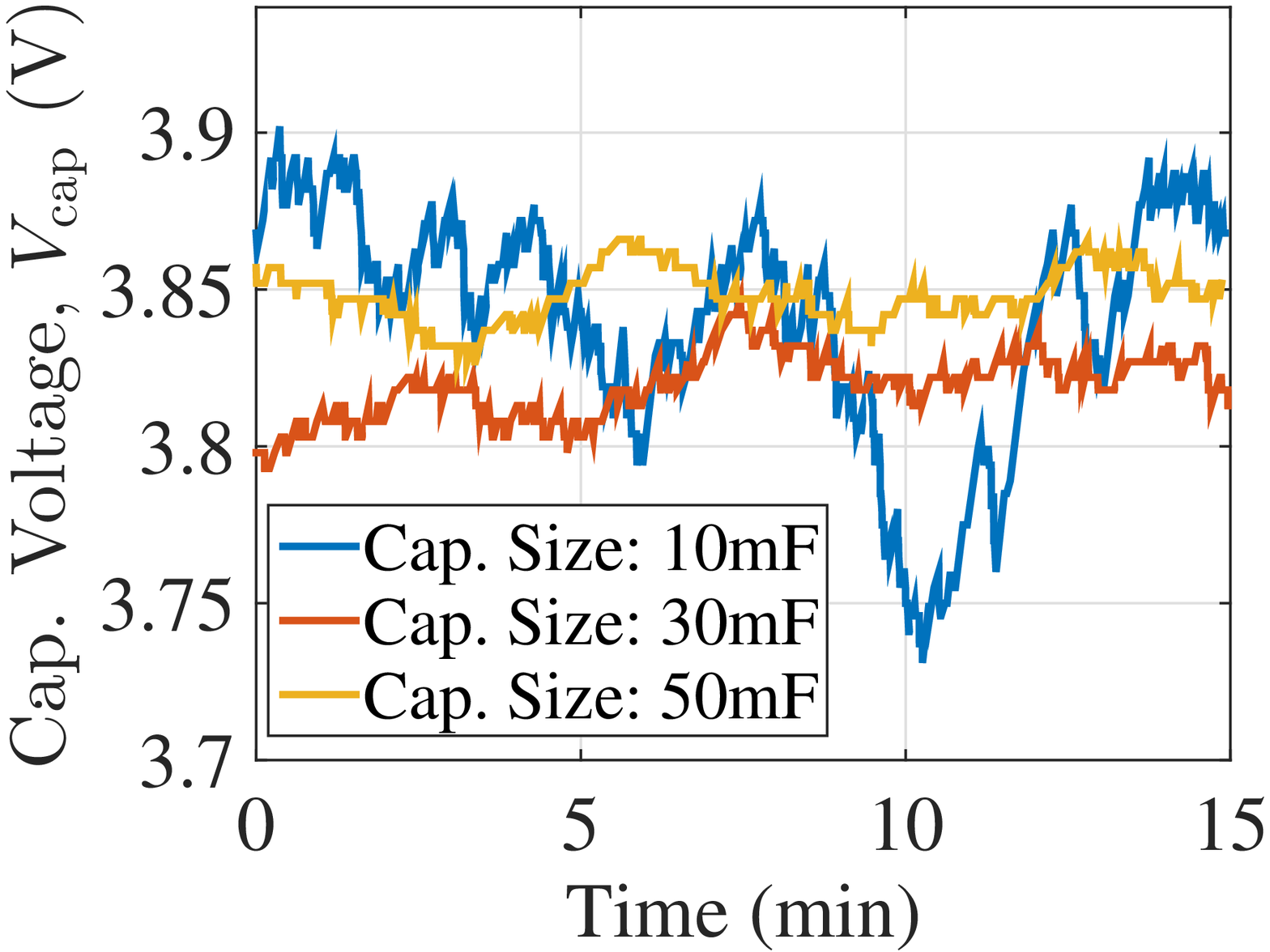}}
\vspace*{-.2in}
\caption{\footnotesize \name experimental evaluation for $N=5$: (a) capacitor voltage level ($V_{\textrm{cap}}$) for a node with a 30\si{\milli\farad} capacitor and $P_b=0.5$\si{\milli\watt}, and (b) capacitor voltage level of nodes with varying capacitor sizes over 15 minutes with $P_b=0.15$\si{\milli\watt}.}
\end{figure}

Using \name, a node consumes power at a rate of up to $P_b$ (\si{\milli\watt}), \emph{on average}. 
 However, the power consumption is stochastic, and therefore, it is expected that the energy stored will vary over time. In Fig.~\ref{fig:voltage}, we show the capacitor voltage over time for a node with $N=5$ and $P_b = 0.5$\si{\milli\watt}. Energy neutrality is demonstrated by the oscillation in the energy level within the limits of the  capacitor storage. 
  Recall from Section~\ref{sec:node_model} that if the energy drains below a software induced threshold of 3.6\si{\volt}, the node temporarily sleeps for 10\si{\second} to regain energy. These periods of additional sleep affect the discovery rate and, as indicated by the accuracy of the experiments, these occurrences are rare.

Furthermore, in Fig.~\ref{fig:varyCap}, we experiment with varying capacitor sizes ranging from 10-50\si{\milli\farad}. As expected, smaller capacitors have added variation in the voltage level. Therefore, smaller capacitors can reach the upper (fully charged) or lower (empty) voltage limits more frequently than larger capacitors. In practice, the capacitor should be sized with respect to the variation in the power consumption and power harvested. 

\subsection{\name Design Considerations}
\label{app:design}
We now consider \name's performance for varying transmit power and discovery message durations.

\subsubsection{Transmit Power, $P_t$} 
\begin{figure}[t]
\centering
\subfigure[\label{fig:sensPt}]{\includegraphics[width=\figSize]{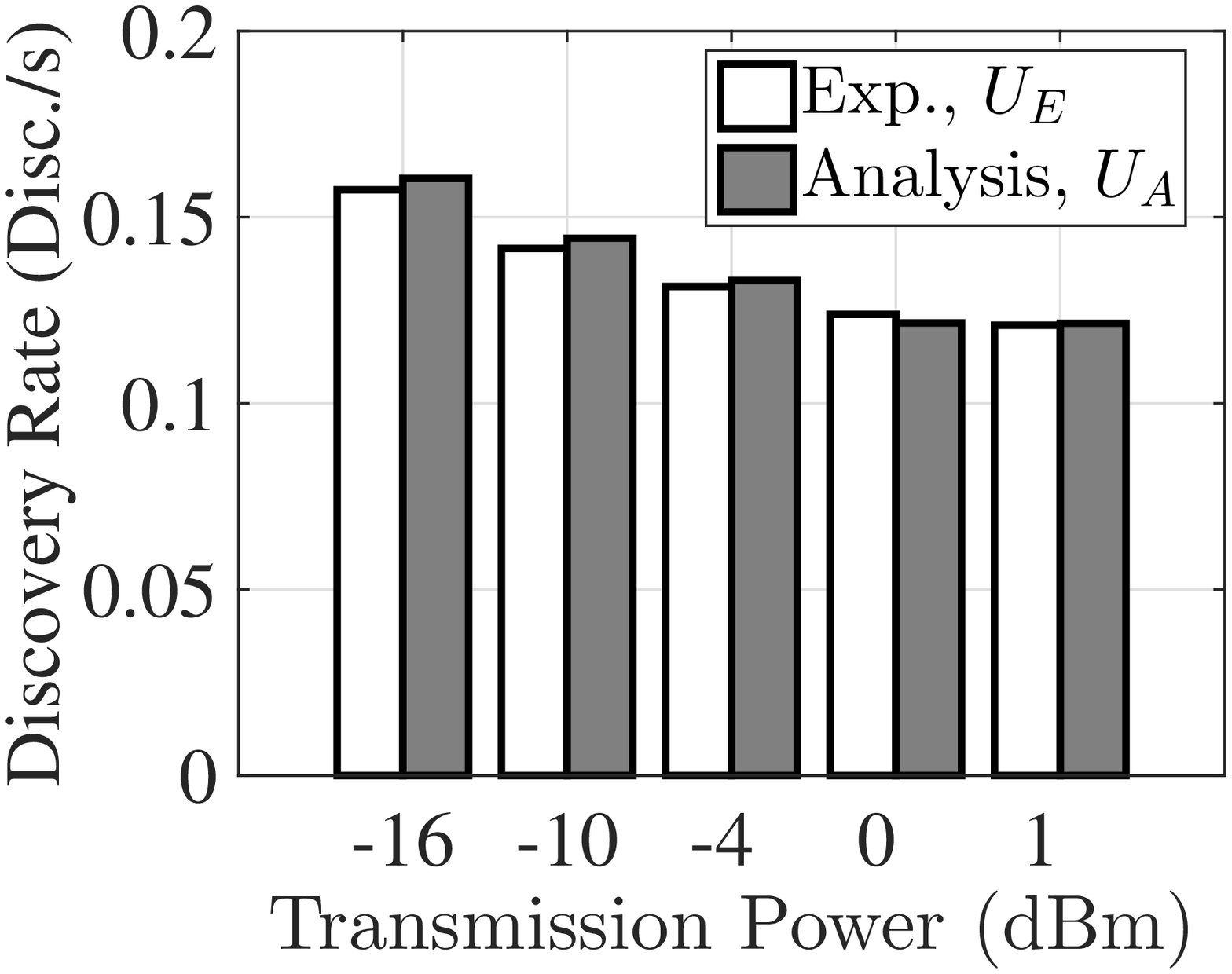}}
\hspace*{0.05\columnwidth}
\subfigure[\label{fig:sensM}]{\includegraphics[width=\figSize]{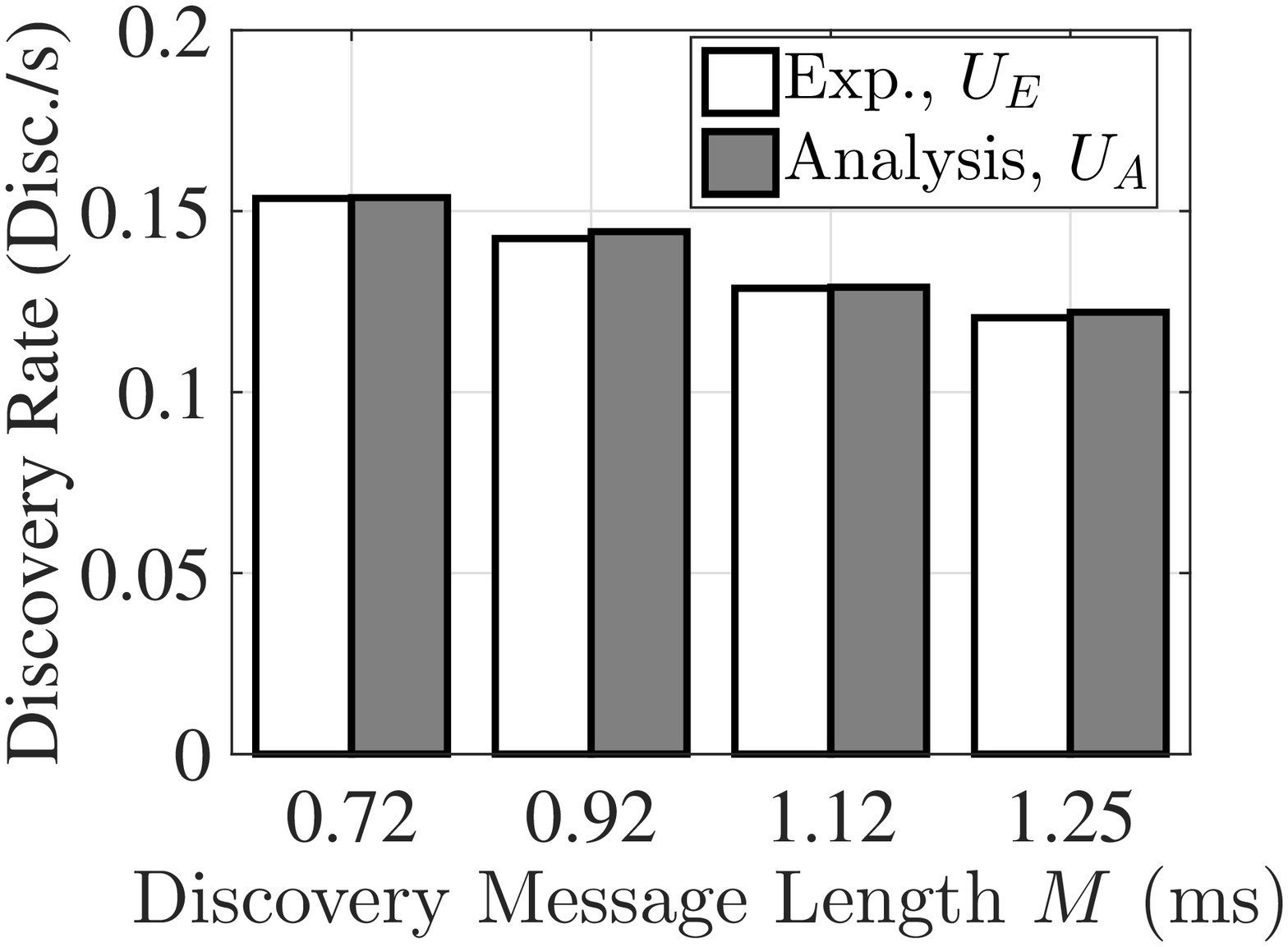}}
\vspace*{-.2in}
\caption{\footnotesize \name performance evaluation for $N=5$ and $P_b=0.5$\si{\milli\watt}: Experimental and analytical discovery rates under varying (a) transmission power ($P_t$) and b) discovery message length ($M$).}
\end{figure}

The transmission power can be set in software. A larger transmission power can result in more geographical coverage, but also consumes more energy. 
In Fig.~\ref{fig:sensPt}, we consider $N=5$ and $P_b = 0.5$\si{\milli\watt} and observe how the discovery rate changes with varying transmission powers. A larger transmission power requires nodes to sleep longer before transmitting, resulting in less discoveries. Note that for this experiment the energy costs from Table~\ref{tab:enercosts} no longer hold and we remeasured them to compute the configuration parameters.

\subsubsection{Discovery Message Duration, $M$} The discovery message requires $M$\si{\milli\second} to be transmitted and contains the node ID and neighbor table information. By adjusting the modulation/coding of the radio or the data content, the packet length can be shortened. A shorter packet length results in less time transmitting as well as less time listening for messages. As shown in Fig.~\ref{fig:sensM}, smaller packet sizes result in an increase in the discovery rate. This presents an application design decision if the contents of the packet can be adjusted to obtain a desired discovery rate. 

\subsection{\name-Dynamic}
\label{sect:practical}
We now evaluate \name-D (described in Section~\ref{sect:dynamic}). The only input to \name-D is the estimated power harvesting rate, $P_b = 0.15$\si{\milli\watt}, and the capacitor voltage $V_{\rm cap}$.  
From~\eqref{eq:blink-d}, the average duration of the exponential sleep is then computed as, 
\begin{align}
\label{eqn:blinkdsleeptime}
	\frac{1}{\lambda} = \frac{382.2238}{V_{\rm cap} - 3.5857} - 2.9843 ~ \textrm{(\si{\milli\second})}.
\end{align}
Thus, the node scales its power consumption based on $V_{\rm cap}$. For example, at $V_{\rm cap} = 3.6$\si{\volt} and 4\si{\volt}, the node will sleep on average for $26.75$ and $0.92$ seconds, respectively. 

To estimate the average sleep duration for a given node in \name-D, 
we compute the average value of $V_{\rm cap}$ over the course of an experiment.
 Based on the this value, the average sleep duration is estimated from~\eqref{eqn:blinkdsleeptime}.

\name-D does not require a priori information of the number of neighbors, $N$. Therefore, throughout this section, (A3) is relaxed. Below, we observe the performance of \mbox{\name-D} first when (i) nodes remain in a clique topology with homogenous power budgets. Then we consider \name-D (ii) in a multihop topology (relaxing (A2)), and finally (iii) in non-homogenous power harvesting scenario (relaxing (A1)). Relaxing all assumptions together requires running a live real-world experiment and is a subject of future work. \comment{Remove last sentence?}

\noindent\emph{(i) Comparison to \name: }
\label{sect:blinkdvblink}
We first evaluate \name-D with an experimental setup similar to the one shown in Fig.~\ref{fig:NDsetup}. Specifically, we consider a network of $N=3$ nodes in close proximity with a power harvesting rate of $P_b=0.15$\si{\milli\watt}.

\begin{figure}[t]
\centering
\subfigure[\label{fig:blinkvsblinkd}]{\includegraphics[width=\figSize]{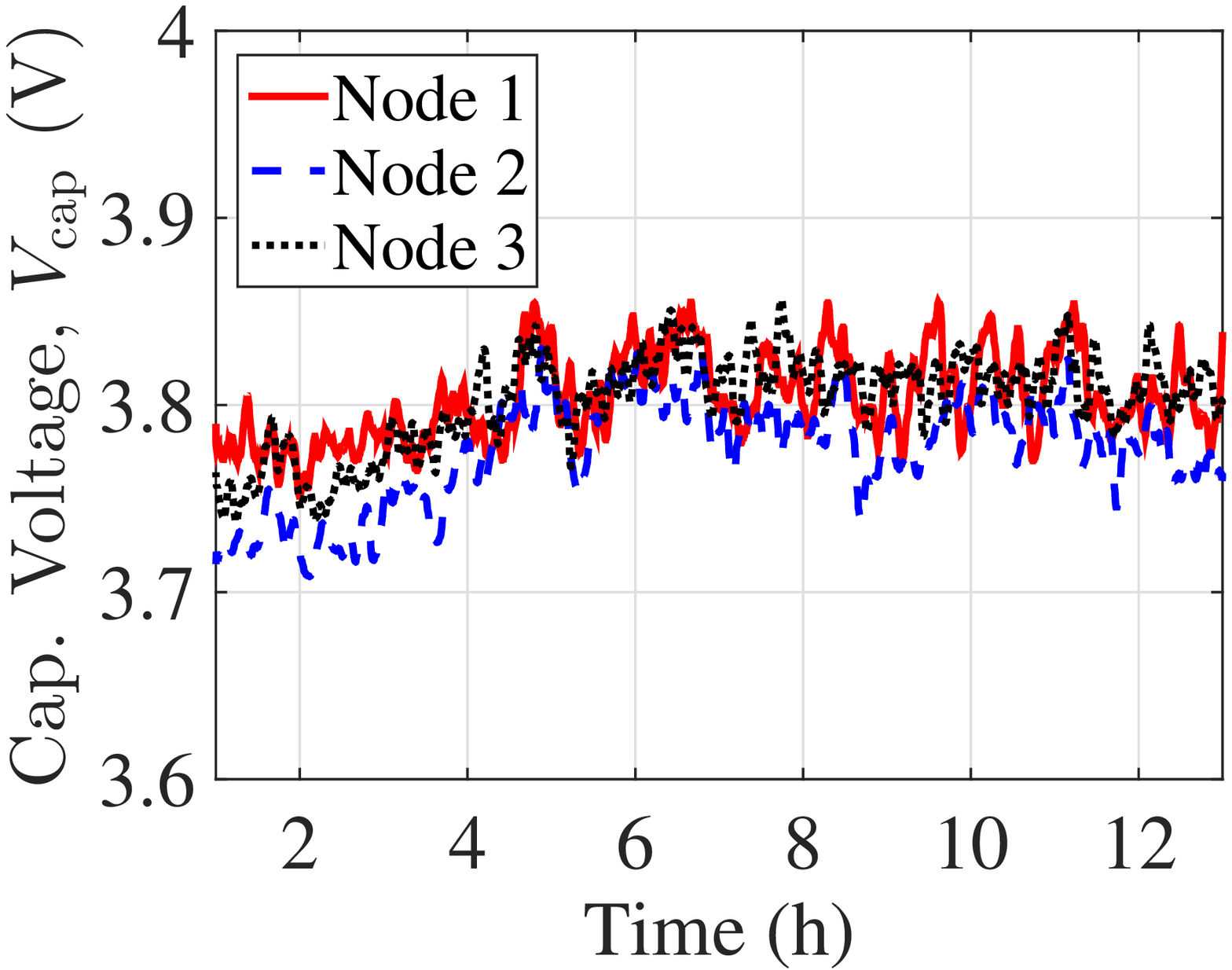}}
\hspace*{0.05\columnwidth}
\subfigure[\label{fig:multihop}]{\includegraphics[width=\figSize]{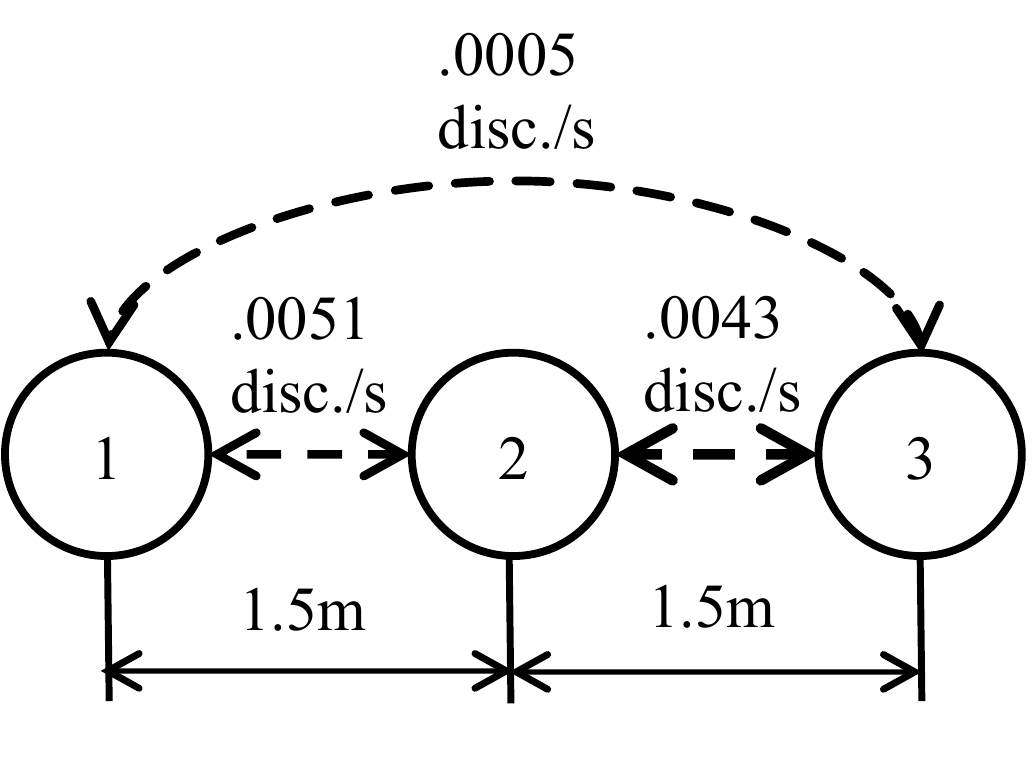}}	
\vspace*{-.2in}
\caption{\footnotesize \name-D experimental evaluation for $N=3$ and homogeneous power harvesting $P_b=0.15$\si{\milli\watt}: (a) Capacitor voltage, $V_{\rm cap}$, in a clique topology. (b) Per link experimental discovery rates for a line topology after $50$ hours.}	
\end{figure}

As shown in Fig.~\ref{fig:blinkvsblinkd}, the capacitor voltage for all 3 nodes stays approximately near $3.8$\si{\volt}. 
As described in Section~\ref{sect:dynamic}, the average power consumption at 3.8\si{\volt} is approximately $P_b$. Therefore, in this scenario, \name-D and \name have similar power consumption and discovery rates. As such, the experimental discovery rate of \name-D is within 1\% of the analytical estimate of \name.

\noindent\emph{(ii) Multihop Topologies: }
Previously, we assumed that all nodes form a clique topology with no packet losses (A2) and the number of nodes $N$ known (A3). Indeed, for the experiments conducted above with a transmission power of \mbox{$-10$dBm}, we found that nodes within $\approx$20m  could be treated as a clique topology with over 99\% packet success rates.

However, to evaluate a non-clique topology and relax (A2) and (A3), we manually reconfigured the transmission power to $-26$dBm and set 3 nodes in a line topology with distance between nodes 1-2 and 2-3 of 1.5\si{\meter}, as shown in Fig.~\ref{fig:multihop}. In this configuration, nodes rarely receive messages from their two-hop neighbors. 
Nodes run \name-D and are given light levels corresponding to the power harvesting rate of $P_b=0.15$\si{\milli\watt} (as described in Section~\ref{sect:testbed}). After 50 hours, the resulting discovery rate is shown on each link in Fig.~\ref{fig:multihop}. 

The two extreme nodes (nodes $1$ and $3$) have very few discoveries from one other, due to the distance between them. However, the node in the middle (node $2$) forms an effective clique of size 2 with each of its neighbors. We therefore can analyze the discovery rate per link. For example, the discovery rate of the link between nodes $1$ and $2$ is $0.0051$ disc./s, 
which is within 1\% of the analytical discovery rate for a clique with $N = 2$ and $P_b = 0.15$mW. Therefore, even with non-clique topologies, each link that is within communication range can be analyzed as a network with $N=2$. This implies that issues such as the hidden-node problem do not significantly affect the performance of \name.

\noindent\emph{(iii) Non-Homogeneous Power Harvesting: }
We now consider nodes 2--5 using \name-D with light levels corresponding to power harvesting of 0.075, 0.15, 0.225, 0.3\si{\milli\watt}, respectively. Node 1 is a \emph{control} node running \name with $P_b=0.15$\si{\milli\watt} and $N=5$.

\begin{figure}
\hspace*{0.1\columnwidth}
\begin{minipage}{\figSize}
\centering
\subfigure[\label{fig:dynamic2}\label{fig:blinkdynamicvoltage}]{\includegraphics[width=0.99\columnwidth]{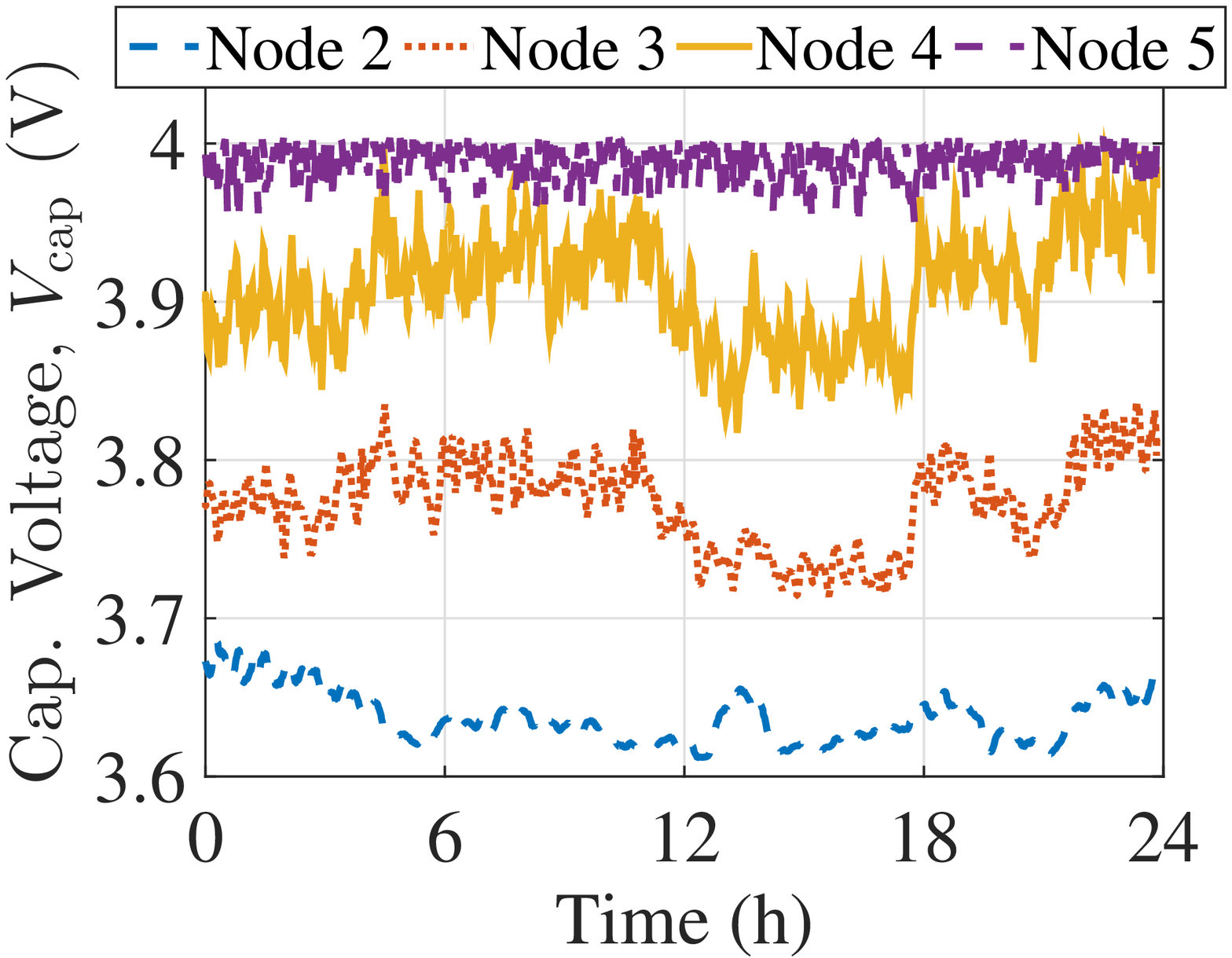}\vspace*{-0.05in}}
\end{minipage}
\subfigure[\label{tbl:dynamictable}]{
\begin{minipage}{0.48\columnwidth}
\vspace*{0.13in}
\centering
\scriptsize
\begin{tabular}{| c | c | c | c | c | c |}
\hline \parbox{0.4in}{\bf \vspace*{0.03in} \centering ND Table\vspace*{0.03in}} & {\bf 1} & {\bf 2} & {\bf 3} & {\bf 4} & {\bf 5} \\ \hline
{\bf 1} & 0  & 11 & 53 & 72 & 96 \\ \hline
{\bf 2} & 9 & 0  & 9 & 15 & 20 \\ \hline
{\bf 3} & 55 & 13 & 0  & 71 & 113 \\ \hline
{\bf 4} & 70 & 16 & 61 & 0  & 177 \\ \hline
{\bf 5} & 93 & 28 & 106 & 175 & 0  \\ \hline
\end{tabular}
\vspace*{.3in}
\end{minipage}
\vspace*{-0.05in}}
\vspace*{-.3in}
\caption{\name-D experimental evaluation for non-homogeneous power harvesting with $N=5$ over 24 hours: (a) Capacitor voltage, $V_{\rm cap}$, and (b) resulting neighbor table. \label{fig:dynamic3}} 
\end{figure}

For each of the 4 \name-D nodes, the capacitor voltage, $V_{\rm cap}$, is shown in Fig.~\ref{fig:dynamic2} and 
settles based on the power harvesting. 
Variations in the settling voltage stem from the dynamic average sleep duration at different power harvesting levels. For example, node 5 is given a light level of 0.3\si{\milli\watt}, and therefore, has a shorter sleep duration than node 2 (light level of 0.075\si{\milli\watt}). Correspondingly, Fig.~\ref{tbl:dynamictable} shows the neighbor table: entry ($i$,$j$) represents the number of discoveries of node $j$ by node $i$ over the experiment duration. Due to non-homogeneity, the discovery rate for each link depends on the power harvested; nodes with larger power budgets discover their neighbors, and are discovered, more frequently.

In Appendix~\ref{app:dynamic_link}, we treat each link with non-homogenous power harvesting as a clique ($N = 2$), and estimate its discovery rate; the approximation is within 20\% of the experimental value.

\section{Conclusions and Future Work}
\label{sect:conc}

We designed, analyzed, and evaluated \name, an ND protocol for EH nodes. By accounting for specific hardware constraints (e.g., transceiver power consumption for transmission, reception, and state switching), \name adheres to a \emph{power budget}. Using renewal theory, we developed the \name Configuration Algorithm (PCA) to determine the nodes' sleep and listen durations which maximize the discovery rate; the PCA achieves a nearly-optimal discovery rate (over 94\%).  

We evaluated \name using TI eZ430-RF2500-SEH EH nodes. The real-life accuracy was consistently within 2\%, demonstrating the practicality of our model. Furthermore, \name outperformed the closest related protocols Searchlight-E~\cite{Bakht_mobicom2012} and BD-E~\cite{McGlynn_mobihoc01} by achieving a discovery rate that was up to 3x higher. 
Finally, we showed that a version of the protocol, \name-Dynamic, was able to adapt to scenarios with non-homogeneous power harvesting and multihop topologies.

\name can be readily applied to nodes with a non-rechargeable battery, where the power budget is set based on the desired lifetime.
Future work will consider relaxing additional assumptions of our model. Primarily, we will attempt to \emph{optimize} \name-D in the presence of nodes with heterogenous power budgets in non-clique topologies. Additionally, we will consider alternate  formulations to achieve closed form optimal configuration parameters.

Finally, we will transform \name into an aggregate-throughput maximizing MAC layer protocol. \name is a natural choice at the MAC layer for applications
requiring information dissemination in infrastructure-less environment 
(i.e., gossip-style routing at the network layer \cite{haas2006gossip}, and data aggregation at the transport layer, such as compressive sensing~\cite{srisooksai2012practical}), as it already maximizes the neighbor discovery rate, which can be easily transformed into a communication rate.



\scriptsize
\bibliographystyle{IEEEtran}
\bibliography{rob,EnergyMAC}

\normalsize
\appendices



\section{Incorporating Idle Power Consumption}
\label{app:sleeping}
In this work, we disregard the power cost of nodes in the sleep state. In this section, we explain how these costs can be incorporated. As described in Section~\ref{sect:experiments}, the idle cost of the microcontroller is normally $1.6$\si{\micro\watt}. This draw is constant for all states (sleep, listen, and idle). As such, to incorporate it into our model, it is simply subtracted from the power budget $P_b$.

In Section~\ref{sect:energy}, we ignore the expected amount of energy (\si{\micro\joule}) consumed by a node when it begins to listen while a packet is currently being transmitted (exemplified by Node $5$ in Fig.~\ref{fig:netRen}). In this case, the node spends energy to transition to and from the sleep state, as well as listen for a short fixed Clear Channel Assessment (CCA) period, denoted as $t_{\rm CCA}$. The energy consumption is then $C_{sl} + P_r \cdot t_{\rm CCA} + C_{ls}$.

This event occurs in a renewal with probability given by $\frac{N-1}{N} (\mathrm{e}^{-\lambda l}) (1 - \mathrm{e}^{-\lambda M})$. Firstly, the node must not be the transmitter w.p. $\frac{N-1}{N}$. As the node is asleep when the transmitter begins to listen, it must then sleep for at least $l$ duration. Finally, given that it is in the sleep state when the transmitter begins to transmit, it must then wakeup before the message is transmitted (duration $M$).

Thus for our experimental evaluation in Section~\ref{sect:experiments}, the idle power costs are summarized in Table~\ref{tab:solutionsSleep} for the experimental parameters originally presented in Table~\ref{tab:solutions}. As can be seen, this probability is very small implying that it is quite rare that a node wakes up in the middle of a transmitted packet. As such, the percentage of the power budget consumed, on average, is always less than $0.5\%$ of the power budget, and therefore can be ignored.  We note, however, that the PCA can easily be modified to incorporate this idle power consumption.

\begin{table}[h!]
\centering
\scriptsize
\caption{\name idle power consumption for every input ($N, P_b$) pair from Table~\ref{tab:solutions}: the probability of a node waking in the middle of a transmitted packet and the expected portion of the power budget ($P_b$) consumed. \label{tab:solutionsSleep}}
\begin{tabular}{| c | c | c | c | c |}
\hline
N & \parbox{1.2cm}{\centering$P_b$ (mW)} & \parbox{4.0cm}{\centering Pr\{A node wakes up in the middle of a packet)\} } & \parbox{2cm}{\centering Expected Energy per Renewal (\si{\micro\joule})} & \parbox{1.2cm}{\centering Pct of $P_b$ ($\%$)} \\ \hline
\multirow{3}{*}{3} & 0.15 & 0.34e-3 & 0.0302 & 0.034 \\ \cline{2-5}
& 0.3 & 0.69e-3 & 0.0605 & 0.068 \\ \cline{2-5}
& 0.5 & 1.15e-3 & 0.1010 & 0.112 \\ \hline
\multirow{3}{*}{5} & 0.15 & 0.41e-3 & 0.0363 & 0.068 \\ \cline{2-5}
& 0.3 & 0.83e-3 & 0.0728 & 0.135 \\ \cline{2-5}
& 0.5 & 1.38e-3 & 0.1215 & 0.223 \\ \hline
\multirow{3}{*}{10} & 0.15 & 0.47e-3 & 0.0410 & 0.151 \\ \cline{2-5}
& 0.3 & 0.94e-3 & 0.0822 & 0.300 \\ \cline{2-5}
& 0.5 & 1.57e-3 & 0.1376 & 0.495 \\ \hline
\end{tabular}
\normalsize
\end{table}



\section{Importance of Switching Costs}
\label{app:switching}

In this work, we incorporate the costs to switch to and from different radio states (sleep, receive, transmit). In this section, we demonstrate the importance of accounting for these costs (which are commonly overlooked in related work).

As the PCA allows for arbitrary switching costs, in Table~\ref{tab:solutionSwitching}, we compute the the parameters assuming that $C_{ij} = 0 \, ( \forall i, j \in \{s, r, t\})$. As indicated in the table, the discovery rate improves by 2-3x compared to the discovery rate when including the switching costs from Table~\ref{tab:enercosts}. However, the power consumed by transition causes the power budget to be exceeded by up to $80\%$. Therefore, ignoring the switching costs may improve the discovery rate, but also results in significantly higher power consumption.

\begin{table}[h!]
\centering
\scriptsize
\caption{\footnotesize \name performance evaluation: the discovery rate $U_A$ resulting from the PCA, and the actual power consumed when ignoring the switching costs.\label{tab:solutionSwitching} }
\begin{tabular}{| c | c | c | c | c |}
\hline
N & \parbox{1.2cm}{\centering $P_b$ (\si{\milli\watt})} & \parbox{4.0cm}{\centering $U_A$, w/ Switching Costs (disc./s)} & \parbox{4.0cm}{\centering $U_A$, w/o Switching Costs (disc./s)} & \parbox{3.6cm}{\centering Power Consumed (\si{\milli\watt})} \\ \hline
\multirow{3}{*}{3} & 0.15 & 0.0039 & 0.010 & 0.26 \\ \cline{2-5}
& 0.3 & 0.0156 & 0.038 & 0.52 \\ \cline{2-5}
& 0.5 & 0.0434 & 0.107 & 0.86 \\ \hline
\multirow{3}{*}{5} & 0.15 & 0.0130 & 0.032 & 0.26 \\ \cline{2-5}
& 0.3 & 0.0519 & 0.128 & 0.52 \\ \cline{2-5}
& 0.5 & 0.1440 & 0.359 & 0.87 \\ \hline
\multirow{3}{*}{10} & 0.15 & 0.0584 & 0.144 & 0.26 \\ \cline{2-5}
& 0.3 & 0.2330 & 0.581 & 0.52 \\ \cline{2-5}
& 0.5 & 0.6470 & 1.630 & 0.87 \\ \hline
\end{tabular}
\normalsize
\end{table}

\section{Software Controlled Light System and Harvesting Inefficiencies}
\label{app:lights}

\begin{figure}[t]
\begin{minipage}{\figSize}
\centering
\includegraphics[width = \columnwidth]{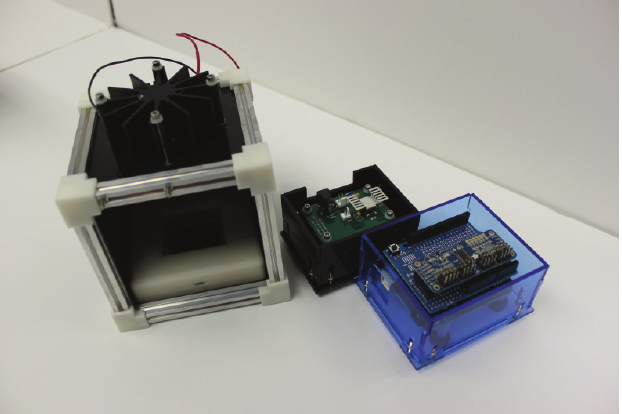}
\caption{\footnotesize Software controlled light system including a dark box enclosure, high-power LED driver, and an Arduino-based light controller.\label{fig:softlight}}
\end{minipage}
\hspace*{0.01\columnwidth}
\begin{minipage}{0.64\columnwidth}
\centering
\subfigure[\label{fig:eh_char_setup}]{\includegraphics[width = 0.49\columnwidth]{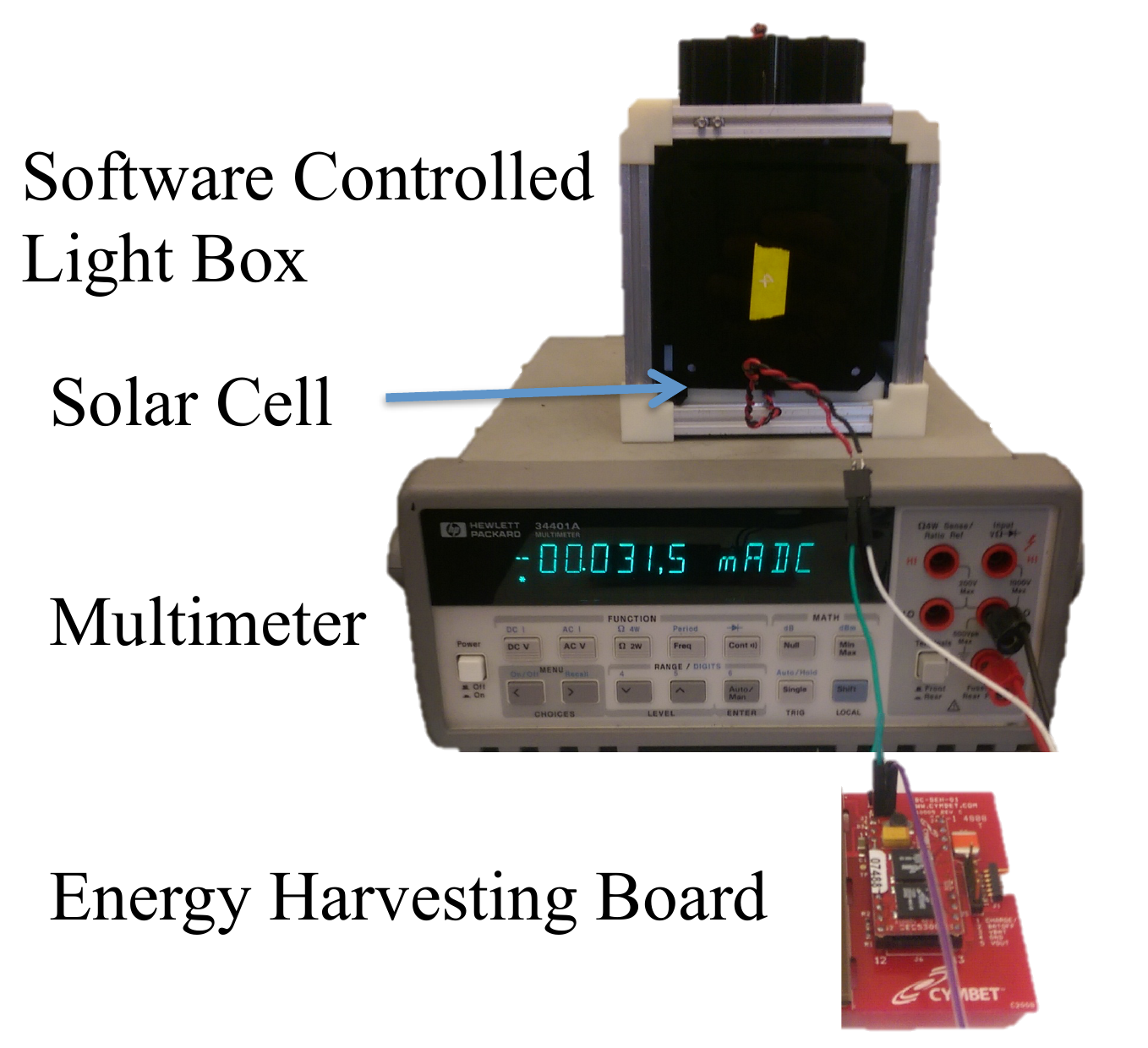}}
\subfigure[\label{fig:eh_char_meas}]{\includegraphics[width = 0.49\columnwidth]{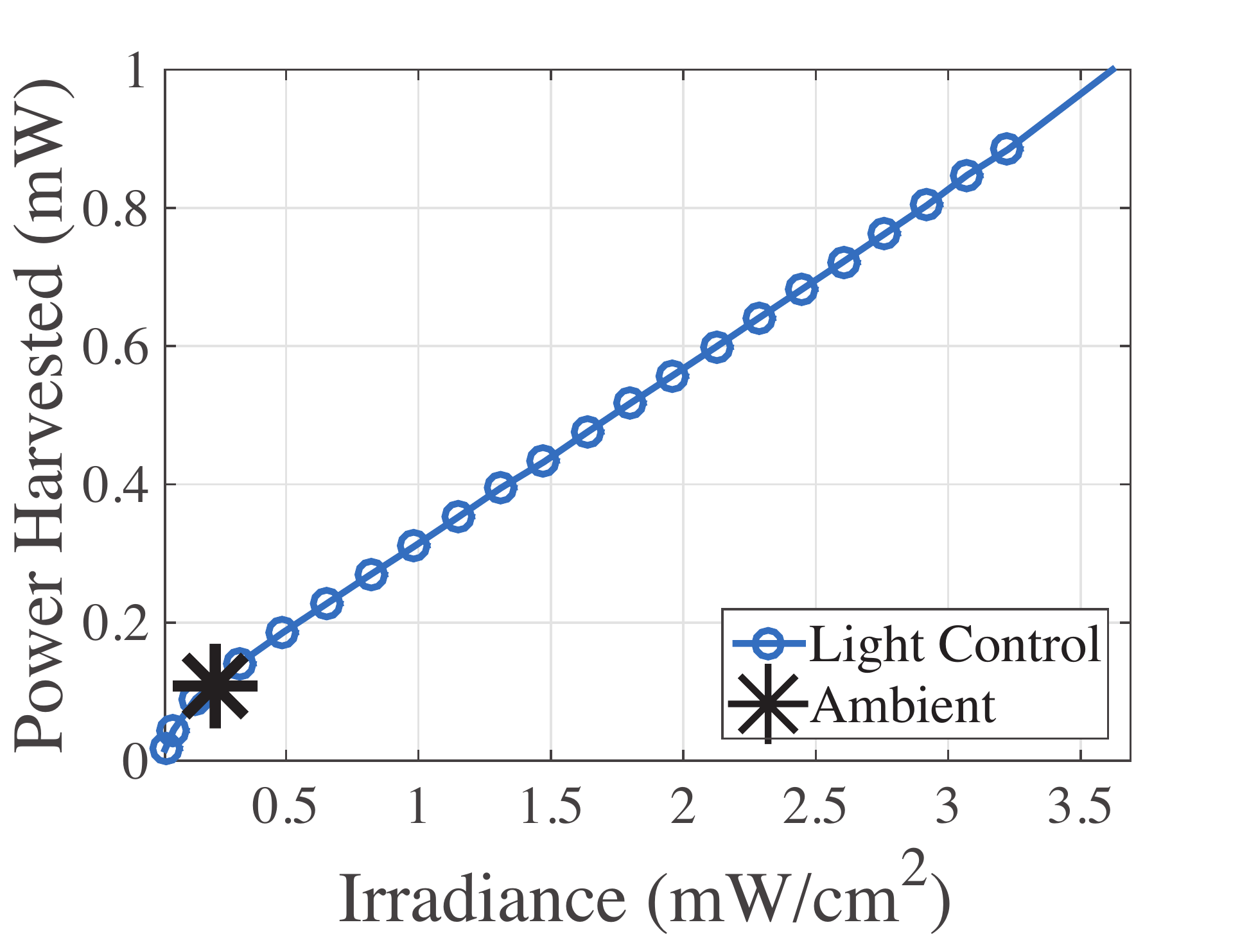}}
\vspace*{-0.1in}
\caption{\footnotesize Energy harvesting characterization: (a) measurement setup and (b) resulting power harvested as a function of the irradiance produced by the software controlled light system.}
\end{minipage}
\end{figure}

We develop an advanced software controlled light system (shown in Fig.~\ref{fig:softlight}) that uses a Java-based script and Arduino-based light control modules to precisely control the irradiance (light energy intensity) generated by LEDs.
The system can produce $1024$ distinct irradiance levels between $0$ and $14$\si{\milli\watt/\centi\square\metre} and the irradiance levels can be changed with time steps of under $100$\si{\milli\second}. Dark box enclosures and 3D printed mounting fixtures ensure full control over the light conditions at the solar cells. This guarantees that our experimental evaluations \emph{are based on the same energy inputs}.

Furthermore, we conduct extensive experiments utilizing a UV818 photodetector to carefully calibrate the irradiance of the light control system. We characterize the power harvested under both the ambient light and the software controlled light setup. The measurement setup is shown in Fig.~\ref{fig:eh_char_setup}. We connect the solar cell in series with a multimeter to measure current. The voltage of the solar cell is at $1.02$V and hence the harvested power can be easily calculated. Then, by sweeping the irradiance using the software controlled light system, Fig.~\ref{fig:eh_char_meas} shows the power harvested for a range of up to 1\si{\milli\watt}.   Using the mapping found in this characterization, we can control the power harvested by the solar cell.

However, the actual power that is stored depends on numerous inefficiencies in the power harvesting circuitry. Specifically, the node contains a Cymbet CBC5300 to up-convert the power harvested from the solar cell at $1.02$\si{\volt} to the capacitor charging voltage of $4$\si{\volt}, which consumes some overhead energy. In addition, there are inefficiencies in the regulation circuit which regulates an output voltage of $3.5$\si{\volt} to power the load. These inefficiencies are difficult to characterize as they vary based on uncontrollable external factors such as the temperature and component variations.

In our evaluations (see Section~\ref{sect:experiments}), nodes were given light levels which corresponded to their power budget $P_b$. Due to the inefficiencies described above, setting the light levels to correspond to each power budget for each node was difficult.

\begin{wrapfigure}{}{3in}
\centering
\includegraphics[width = \figSize]{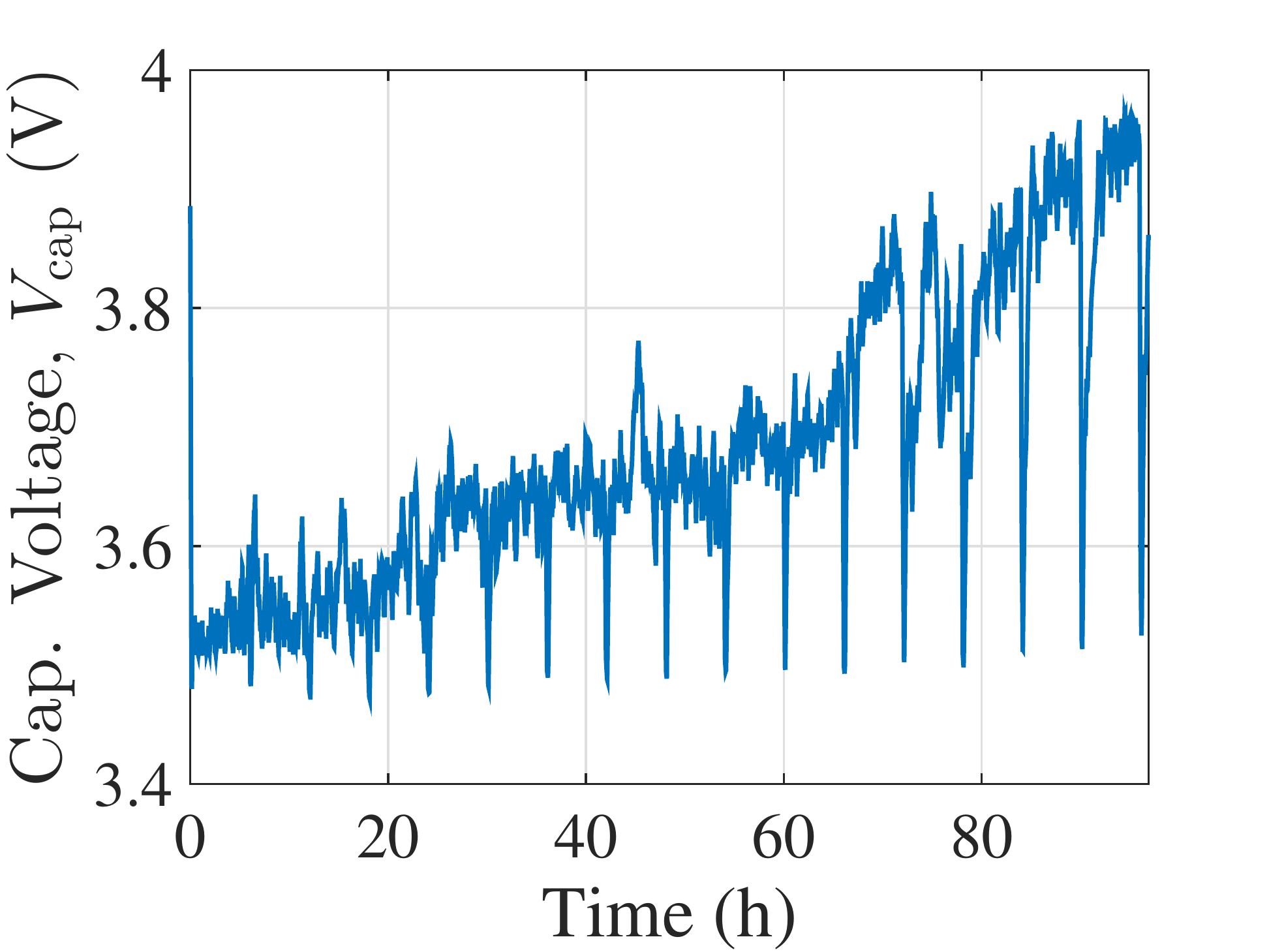}
\vspace*{-.1in}
\caption{A four day experiment to find the light level where the node stays around its energy neutral point.\label{fig:find_neutral_point}}
\end{wrapfigure}

To accomplish this, we conduct a $4$-day experiment in which nodes operated using \name, yet we varied the light levels every $6$ hours. An example of the capacitor voltage for one node in this experiment is shown in Fig.~{\ref{fig:find_neutral_point}}. Each valley represents a $10$-minute ``dark'' period where the light is completely off before changing to the next light levels. 

With limited light levels (i.e., hours $0$-$20$ in Fig.~\ref{fig:find_neutral_point}), the capacitor voltage operates near the minimum implying that the node is consuming more energy than it harvests. With larger light levels (i.e., hours $80$-$100$), the node is harvesting more energy than it consumes and thus the capacitor voltage reaches its upper limit. However for the range of lights corresponding to $20$-$80$ hours in Fig.~{\ref{fig:find_neutral_point}}, the node has a relatively stable voltage, implying that it is consuming power (on average) as the same rate it harvests; \emph{energy neutrality} is obtained.

By performing this experiment for all nodes, we found the light levels at which each node is energy neutral. The  neutral light levels varied significantly. Furthermore, by comparing the power harvested by the solar cell to the power budget ($P_b$), we found that the efficiency of the storage process to be between $40\%$ and $60\%$. This emphasizes the need to incorporate energy storage feedback into the ND protocol, as is done by Panda-D.


\section{Description of Searchlight-E and BD-E}
\label{app:related}
In this section, we describe our adaptations to two related protocols, termed Searchlight-E~\cite{Bakht_mobicom2012} and Birthday-E (BD-E)~\cite{McGlynn_mobihoc01}. Both of these protocols are based on nodes maintaining time slots. We will denote the time slot duration as $d_s$. In each \emph{active} time slot, a node sends a beacon message (originally proposed by~\cite{Dutta_disco08}). The beacon begins a slot with a packet transmission, then listens to the channel, and ends the slot with a packet transmission.

In the Searchlight protocol, two slots are active per cycle of length $t$ slots. Therefore, the power budget can be written as
\begin{align*}
\frac{2 \left(2 P_t M + P_r \left( d_s - 2M \right) + C_{st} + C_{ts} \right)}{t d_s} \leq P_b.
\end{align*}
In the Birthday protocol, each node transmits a beacon in a slot with probability $p$. Therefore, the power budget is simply
\begin{align*}
p \left(2 P_t M + P_r \left( d_s - 2M \right) + C_{st} + C_{ts} \right) \leq P_b.
\end{align*}
In our evaluation, we select $t$ and $p$ such that the power budget is fully consumed and term these protocols, Searchlight-E and BD-E.

We note that there are numerous aspects of related protocols~\cite{McGlynn_mobihoc01, Dutta_disco08, Bakht_mobicom2012, sun2014hello, Kandhalu_uconnect10} which have not been considered. Specifically, existing works do not consider collisions occurring due to no clear channel assessment.\footnote{ \name avoids collisions by always listening before transmitting.} Furthermore, numerous practical parameters are not considered such as the setting of the slot size. As can be seen above, the slot size impacts the average power consumption. In our simulation of Searchlight-E and BD-E, we ignore  collisions and set the slot size to $d_s = 50$\si{\milli\second} with a guard time of $1$\si{\milli\second}, as was done in~\cite{sun2014hello}.

\section{Approximate Analysis of \name-D}
\label{app:dynamic_link}
We now show how to compute the approximated directional discovery rate $U_{ij}$ (i.e., the rate at which node $i$ is discovered by node $j$) for a link between node $i$ and $j$, under non-homogeneous power budgets $\lambda_i$ and $\lambda_j$, respectively. Using similar analysis as in Section~\ref{sec:protocol}, we obtain
\begin{align}
\label{eq:non_homo_link_disc_rate}
U_{ij} = \frac{\frac{\lambda_i}{\lambda_i + \lambda_j} (1 - e^{-\lambda_j l})}{\frac{1}{\lambda_i + \lambda_j} + l + M},
\end{align}
in which $\frac{\lambda_i}{\lambda_i + \lambda_j}$ is the probability that node $i$ becomes the transmitter in a renewal. 
\begin{wrapfigure}{}{4in}
\centering
\scriptsize
\captionof{table}{\name-D Experimental and Predicted Parameters for the $N = 5$ experiment from Fig.~\ref{fig:dynamic3}. 
 \label{tab:blink-d-param}}
\begin{tabular}{| c | c | c | c | c | c |}
\hline
& {\bf 1} & {\bf 2} & {\bf 3} & {\bf 4} & {\bf 5} \\ \hline
{\rm Exp. Avg. Voltage (V)} & 3.5142  & 3.6432 & 3.7772 & 3.9122 & 3.9880 \\ \hline
{\rm Exp. Avg. Sleep Duration, $1/\lambda_i$ (ms)} & 1777.2 & 7542.6  & 2041.2 & 1181.5 & 947.9 \\ \hline
{\rm Est. Avg. Sleep Duration, $1/\lambda_i$ (ms)} & 1777.2 & 6644.4  & 1993.0 & 1167.7 & 947.1 \\ \hline
\end{tabular}
\normalsize
\end{wrapfigure}
To evaluate this approximation, we apply it to the non-homogenous power harvesting experiment presented in Fig.~\ref{fig:dynamic3}. Recall that the sleep rate for a node $i$, $\lambda_i$, is dynamically changing in \name-D. We estimate the sleep rate based on the experimental average capacitor voltage using \eqref{eq:blink-d}, which is shown in Table~{\ref{tab:blink-d-param}}.

\begin{wrapfigure}{}{3in}
\centering
\scriptsize
\captionof{table}{\name-D Discovery Rate Approximation for the $N = 5$ experiment from Fig.~\ref{fig:dynamic3}: Error rate (\%) of the experimental per-link discovery rate compared to \eqref{eq:non_homo_link_disc_rate}. \label{tab:blink-d}}
\begin{tabular}{| c | c | c | c | c | c |}
\hline
 & {\bf 1} & {\bf 2} & {\bf 3} & {\bf 4} & {\bf 5} \\ \hline
{\bf 1} & -   & -17 & 8   & -15 & -9  \\ \hline
{\bf 2} & -32 & -   & -22 & -23 & -20 \\ \hline
{\bf 3} & 12  & 12  & -   & -4  & 23  \\ \hline
{\bf 4} & -17 & -20 & -17 & -   & 12  \\ \hline
{\bf 5} & -11 & 13  & 16  & 11  & -   \\ \hline
\end{tabular}
\normalsize
\end{wrapfigure}

Corresponding to the parameters in Table~\ref{tab:blink-d-param}, in Table~\ref{tab:blink-d} we compute the error rate between the experiment per-link discovery rate and~\eqref{eq:non_homo_link_disc_rate}. In general, the approximation is quite crude (typically within 25\%). Yet it can still be used as a rough approximation of the per-link discovery rate.
We remark here that the relatively high errors come from: (1) the small number of discoveries, (2) the fact that each node is performing independently without knowing the the value of $N$ as a-priori, and (3) errors in the ADC capacitor voltage sampling. We observe that even after relaxing assumptions (A1, A2, A3), \name-D still has robust performance in terms of the per-link discovery rate.


\end{document}